\documentclass[preprint,review,12pt]{elsarticle}%review,%[preprint,12pt]%final,1p,times
\usepackage{enumitem}
\usepackage{bm}
\usepackage{url}
\usepackage{caption}
\usepackage{amsmath}
\usepackage{graphics}
\usepackage{amssymb}
\usepackage{color}
\journal{Journal of sound and vibration}

\begin{document}

\begin{frontmatter}
\title{New algorithm for footstep localization using seismic sensors in an indoor environment}
\author{R. Bahroun$^1$, O. Michel$^2$, F. Frassati$^1$, M. Carmona$^1$, J.L. Lacoume$^{1,2}$}
\address{$^1$ CEA-Leti, MINATEC-Campus. 17 avenue des Martyrs, 38054 Grenoble, Cedex 9, France,}
\address{$^2$Gipsa-lab, Grenoble Universit\' e. 961 rue de la Houille Blanche, BP 46 F- 38402 Grenoble Cedex, France}

\begin{abstract}
In this study, we consider the use of seismic sensors for footstep localization in indoor environments. A popular strategy of localization is to use the measured differences in arrival times of source signals at multiple pairs of receivers. In the literature, most algorithms that are based on time differences of arrival (TDOA) assume that the propagation velocity is a constant as a function of the source position, which is valid for air propagation or even for narrow band signals. However a bounded medium such as a concrete slab (encountered in indoor environement) is usually dispersive and damped. In this study, we  demonstrate that under such conditions,  the concrete slab can be assimilated to a thin plate; considering a Kelvin-Voigt damping model, we introduce the notion of {\em perceived propagation velocity}, which decreases when the source-sensor distance increases. This peculiar behaviour precludes any possibility to rely on existing  localization methods in indoor environment. Therefore, a new localization algorithm that is adapted to a damped and dispersive medium is proposed, using only on the sign of the measured TDOA (SO-TDOA). A simulation and some experimental results are included, to define the performance of this SO-TDOA algorithm. 

\end{abstract}

\end{frontmatter}
%---------------------------
\section{Introduction}
For many applications, it is important to obtain location information about a resident in an indoor environment. For example, knowing the position of a resident can facilitate the control of the heating and air conditioning systems. Existing solutions, however, are intrusive, and they do not respect the private life of the resident (e.g., audio or video monitoring \cite{Ali_2006}), or they are obliging people to keep sensor on their body all the time (e.g., the magneto-inertial navigation technique \cite{MINAV_2012}). In this study, we propose a new indoor localization algorithm that is not constrained. This new algorithm is based on seismic signal processing.\\
The vibration signature of the human footstep on a floor creates an elastic wave that is induced by the walking motions. Our goal is to localize footsteps using seismic sensors that are fixed on the floor in the indoor environment. Only a few studies have described seismic methods that are applicable to footstep localization in an indoor environment. The present techniques can be divided into two groups:
\begin{itemize}
\setlist{nolistsep}
\item Techniques based on seismic-wave structures \cite{Heyoung_2009, Stafsudd_CMA_SWA_2007}: with this type of technique, a footstep is modeled as a seismic signal composed of P-waves (longitudinal waves) and S-waves (transversal waves) in a three-dimensional environment. Using this assumption, the direction of arrival can be determined from the correlation between the signals recorded by a three-axis accelerometer. These techniques which where initially devised for outdoor envoironments cannot be easily transposed in indoor environments. Indeed, the signals recorded indoors by a sensor is a mix of direct and reflected waves (e.g., reflections on the edge of the slab, reflections on the furniture and facilities) in an almost two-dimensional environment. A concrete slab in a building is better modeled by a thin plate than by a semi infinite half space propagation medium. Propagating flexural waves dominate the response.\\    
The time delay between two paths is very short in an indoor environment. The distances are only a few meters and the propagation velocity of seismic waves is more than $1000 \mathrm{m/s}$ in a concrete medium. In addition, elastic waves propagated on the floor depend on many factors; among these, the footwear of the person, the angle of impact excitation, the construction of the floor, and the geometrical walking pattern \cite{Persson_2008, Ekimov_2006, Bates_1983} are important factors, among others. The physical characteristics of the medium itself (concrete) exhibit a high variablity, with an important impact on the wave propagation velocity wich may vary from one sector to another on the same slab. As a consequence, cross-correlation based approaches specifically derived for source location in thin plates (see e.g. \cite{Ziola_1991}) cannot be used here.      
\item Techniques based on range delay estimation \cite{Richman_2001, Zheng_2007}: these techniques, such as hyperbolic localization \cite{Chan_1994}, are based on time differences of arrival (TDOA) and the propagation velocity estimation. The propagation velocity is assumed to be constant and independent of the source position. In other words, the time of arrival (TOA) depends linearly on :sensor distance. 
\end{itemize}

In what follows, we will first discuss the applicability of localization techniques assuming a constant propagation velocity for the problem of footstep localization using seismic sensors. Indeed, because the various wave components travel at different propagation velocities, footstep signals will vary from one receiver location to another. The detected arrival times and the perceived propagation velocities will closely depend on the attenuation and the dispersion properties of the floor. A theoretical study of elastic-wave propagation based on a simplified bending-wave equation will be conducted in section \ref{TDOA}. This study will show that the perceived propagation velocity decreases in a floor assimilated to a thin, damped, and dispersive plate if the source-sensor distance increases. Analytical and experimental results will also be presented to reinforce this conclusion. Therefore localization techniques based on range delay estimation are inadequate for our problem.\\
A new localization algorithm will be proposed in section \ref{Principle_SO-TDOA}. This new algorithm takes into account the nonconstant propagation velocity and exploits the property that the order of arrival of the signals at the sensors is maintained in the dispersive and damped floor being considered. The proposed footstep localization algorithm is based on a study of the sign of the time differences of the arrival (SO-TDOA). The development of the proposed algorithm will be followed in section \ref{perf}, where we describe simulation results and analyze the performances of the proposed SO-TDOA algorithm, as compared with the hyperbolic algorithm that is based on range estimation. Section \ref{exp} will describe the field tests and provide some experimental results. 
%%-----------------
\section{Perceived propagation velocity of the seismic signal of a footstep on a floor}
\label{TDOA}
The floor of an indoor environment will be assimilated to a thin damped isotropic plate \cite{Allen_1998, Thomas_2003} throughout this study. Considering this assumption, the goal of this section is to define the influence of the dispersion and the damping effects on the "perceived propagation velocity" estimated by a given measuring strategy. \\
Consider a plate of thickness $h$, of infinite extent in the $x,y$ plane. The governing equation for the bending motion of a thin undamped plate is \cite{Graff_1975, Onde_livre}:
\begin{equation}
\rho h \frac{\partial^2}{\partial t^2} u(x,y,t)+ D \triangle^2 u(x,y,t) = f  
\label{ss_disp}
\end{equation} 
where $u$ is the transversal displacement, $D = E \frac{h^3}{12(1-\sigma^2)}$ is the bending stiffness, $E$ is the Young's modulus, $\sigma$ is the Poisson ratio, $\rho$ is the mass density, $\triangle = \frac{\partial^2}{\partial x^2} + \frac{\partial^2}{\partial y^2}$ is the Laplacian, and $f$ describes the external forces exerted on the plate. Eq. (\ref{ss_disp}) corresponds to the ordinary flexural wave equation. It is satisfied for a thin plate where its thickness $h$ is less than a sixth of the wavelength ($h <\lambda/6$). A correction term can also be added in the case of a thick plate, to represent the effects of shear stress (although this is not the case in the present study). \\ 
Internal mechanical damping is taken into account by introducing a viscous friction force. This friction force is proportional to the time derivative of the strain. Thus Eq. (\ref{ss_disp}) for a damped medium is given by the Kelvin-Voigt model \cite{DR_1996, Heckl_2010}:
\begin{equation}
\rho h \frac{\partial^2}{\partial t^2} u(x,y,t)+ D \left(1+ \vartheta \frac{\partial}{\partial t} \right) \triangle^2 u(x,y,t) = f 
\label{avec_disp}
\end{equation} 
Then, the dispersion relation is deduced:
\begin{equation}
-\omega^2+ a^2 \left(1- j\vartheta \omega \right) k^4 =0
\label{eq3}
\end{equation}
where $\eta = \vartheta \omega$ is the dimensionless loss factor that is characteristic of the damping effect, and $a$ is a characteristic of the concrete slab, such that $a = \sqrt{\frac{D}{\rho h}}$ $[\mathrm{m^2s^{-1}}]$. This implies that:
\begin{equation}
k(\omega)=\sqrt{\frac{\omega}{a}} \left( 1 - j\vartheta \omega \right)^{-\frac{1}{4}}
\label{eq4}
\end{equation}
for a low loss factor ($\vartheta \omega << 1$),
\begin{equation}
k(\omega) \simeq \sqrt{\frac{\omega}{a}} \left( 1 + \frac{1}{4}j \vartheta \omega \right)=k_R(\omega)+jk_I(\omega)
\label{eq5}
\end{equation}
where $k_R$ and $k_I$ are the real and the imaginary parts of the wave number $k$, respectively. 
$k_I$ is known as the attenuation coefficient of the wave in the propagation direction. So the damping induces frequency-dependent attenuation ($k_I(\omega)$). The dispersion ($k_R(\omega)$) causes a frequency-dependent group velocity propagation $c_g$ that is given by:
\begin{equation}
c_g=\frac{\partial \omega}{\partial k_R}\simeq 2 \sqrt{a} \sqrt{\omega}. 
\label{speed} % eq6
\end{equation} 
Considering the hypothesis of a low loss factor (cf. Eq. (\ref{eq4}) and Eq. (\ref{eq5})), an approximate expansion of the propagating wave packet as a Fourier integral is proposed below. Detailed calculation and explanations are given in  \ref{annexeB} (Note that the derivations in the appendix are presented for a 1D case only, for sake of feasibility and are assumed to hold in the present 2D case). 
If the propagation medium is isotropic, the transversal displacement $u(x,y,t)$ depends only on the source-sensor distances $d$. For an initial wave at position $(x,y) = (0,0)$, the propagative wave $u(d,t)$ at another position of distant of $d$ is given by: 
\begin{eqnarray}
u(d,t)&\propto& \int U(0,\omega) \mathrm{e}^{j[k(\omega) d- \omega t]} \mathrm{d }\omega \\
&=& \int U(0,\omega) \mathrm{e}^{-k_I(\omega) d} \mathrm{e}^{j[k_R(\omega) d- \omega t]} \mathrm{d} \omega
\label{eq7}
\end{eqnarray}
where $U(0,\omega)$ is the spectrum of the wave $u(0,t)$.\\
Using the plate bending wave equation, we derived the  dispersion relation in Eq. (\ref{eq3})-(\ref{eq5}) under the assumption of a low loss factor. Then we deduced the group velocity in Eq. (\ref{speed}). However,  group velocity is not sufficient by itself to model the time of arrival of a wide band wave packet in dispersive and dissipative media: the spectral content of the wave packet evolves during propagation, as attenuation occurs. Consequently the dominant frequency of the packet decreases as the wave packet propagates; the group velocity estimated from Eq. (\ref{speed}) at the central frequency of the wave packet decreases as this latter central frequency decreases (dissipation occurs mainly on the high frequency part of the spectrum). This motivates the introduction of the heuristic notion of "perceived propagation velocity", which is simply related to the estimated time of arrival of the wave packet.\\
The purpose of the next section is to study the variation of this "perceived propagation velocity" with the source-sensor distance, for a given shape of the excitation term $f(x,t)$\footnote{An adequate choice of $f(x,t)$ turns out to be crucial for insuring convergence of the integrals in Eq (\ref{eq7}), or simply to allow analytical derivations. This is discussed in full details in  \ref{annexeB}}.    
Two approaches are presented. The first one consists in evaluating the integral in Eq. (\ref{eq7}) using a discrete sum, and then applying a threshold to detect the time of arrival. Thus it provides an estimate of the  "perceived propagation velocity" for a given distance. The second approach consists in using the stationary phase method to evaluate the envelope of the signal in Eq. (\ref{eq7}). Then the relationship between the "perceived propagation velocity" and the source-sensor distance can also be derived. \\
We consider a concrete slab of thickness $h = 20 \mathrm{cm}$, Young's modulus $E = 24. 10^{-9} \mathrm{N/m^2}$, mass density $\rho = 2500 \mathrm{kg/m^3}$, and Poisson's ratio $\sigma = 0.2$, \cite{Heckl_2010}. Under such conditions, $a$ is about $183 \mathrm{m^2/s}$. This indicated value is useful as an example, because the mechanical properties of a material like concrete are known to depend strongly on their composition and how they are made. It is also important to note that the expression in Eq. (\ref{eq7}) is not valid at short source-sensor distances (e.g., $\sim 1 \mathrm{m}$) considering a plate of thickness $h\sim 20 \mathrm{cm}$. Indeed, the approximation of a thin plate ($h <\lambda/6$) is not valid at these distances because the signal is dominated by high frequency components, which is equivalent to a short wavelength ($\lambda <1 \mathrm{m}$).\\
In the literature, the loss factor of concrete material $(\eta = \vartheta \omega)$ can take values from $10^{-3}$ to $10^{-2}$ in the audio frequency range \cite{Heckl_2010}. Without the lose of generality, we choose $\vartheta = 10^{-5}\mathrm{s}$ for $\omega <10^4 \mathrm{Hz}$. The approximation $(\vartheta \omega <<1)$ is then satisfied for the concrete medium.\\
In the sequel, it will be assumed that the choice of $f(x,t)$ leads to 
$U(0,\omega_0) = \alpha \omega^{-3/2} \hat{f} \simeq \alpha \omega^{1/2}$ where $\alpha = \frac{j\pi a^{3/2}}{2D}$. Refer to  \ref{annexeB} for details. 

%-----------
\subsection{Perceived propagation velocity - integral approximation}
\label{Sim}
To simulate the received signal $u(d,t)$ at a distance $d$ from the source, an approximation of the infinite integral in Eq. (\ref{eq7}) using a discrete finite sum is proposed, with:
\begin{equation}
u(d,t) \simeq \frac{\omega_m}{n} |\alpha| \sum_{i=0}^{n-1} \left( \frac{i\omega_{m}}{n} \right)^{1/2} \mathrm{e}^{-k_I\left( \frac{i\omega_{m}}{n} \right) d} \mathrm{cos} \left( k_R \left( \frac{i\omega_{m}}{n} \right) d- \left( \frac{i\omega_{m}}{n} \right) t+\pi/4 \right) 
\label{udt}
\end{equation}
where $\omega_{m} = 2 \pi f_{\max}$. $f_{\max}$ is fixed at $10 \mathrm{kHz}$ and $n = 2048$ for this simulation. \\
It should be noted that the approximation in Eq. (\ref{udt}) is not valid for short source-sensor distances ($d <3 \mathrm{m}$), because at these distances the signal is dominated by high frequency components that are not considered by the finite sum in Eq. (\ref{udt}).\\

\begin{figure}[!ht]
\centering
\includegraphics[width=12cm]{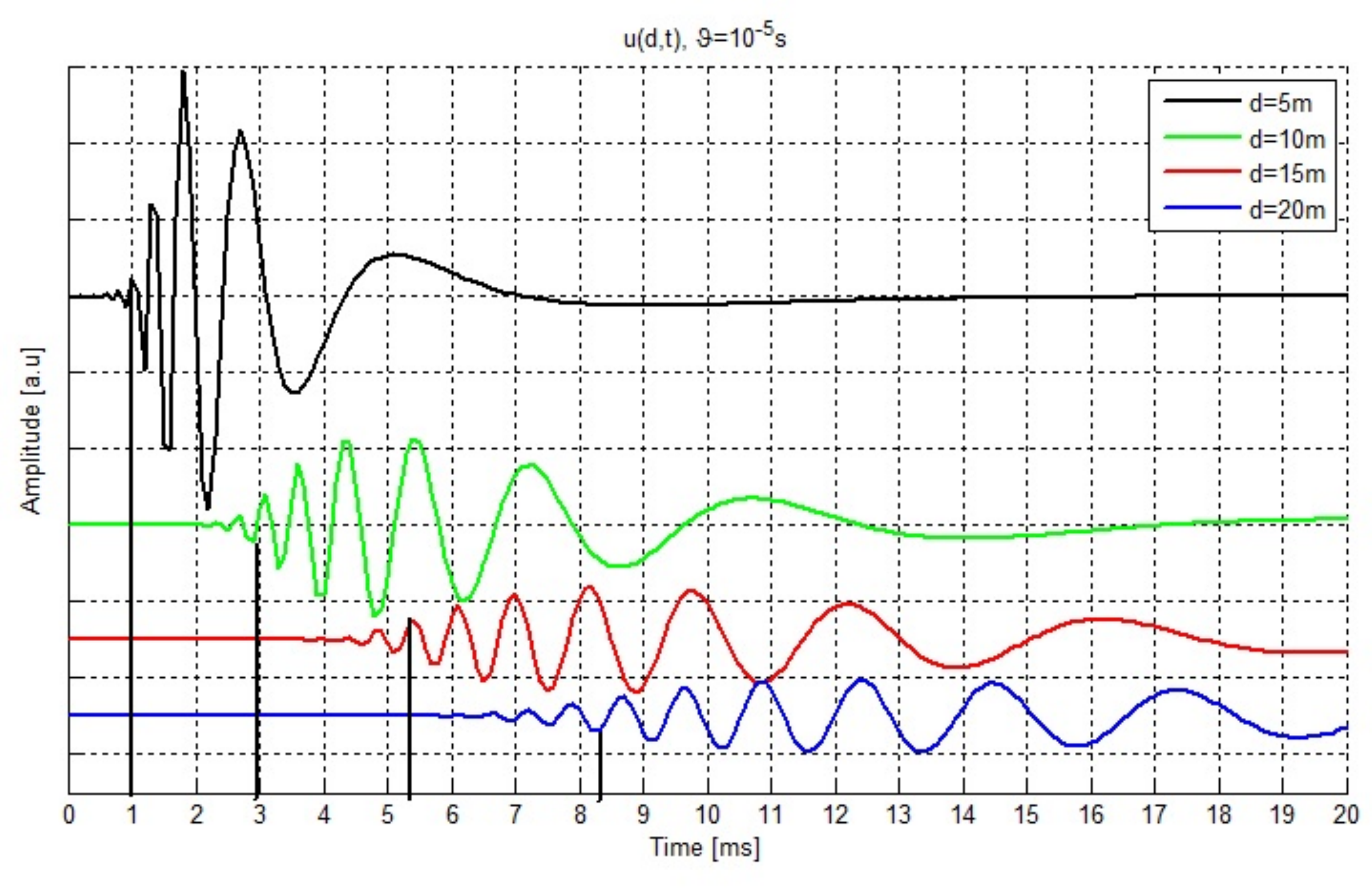}%d5101520_dec2.pdf}
\centering
\caption{\small{Simulated received signal at $d=5,10,15,20 \mathrm{m}$.}}
\label{d5101520}
\end{figure}

Figure \ref{d5101520} shows the simulated signals received at $d = 5, 10, 15, 20 \mathrm{m}$. The amplitude scale is in arbitrary units.\\

\begin{figure}[h!]
\centering
\includegraphics[width=14.5cm]{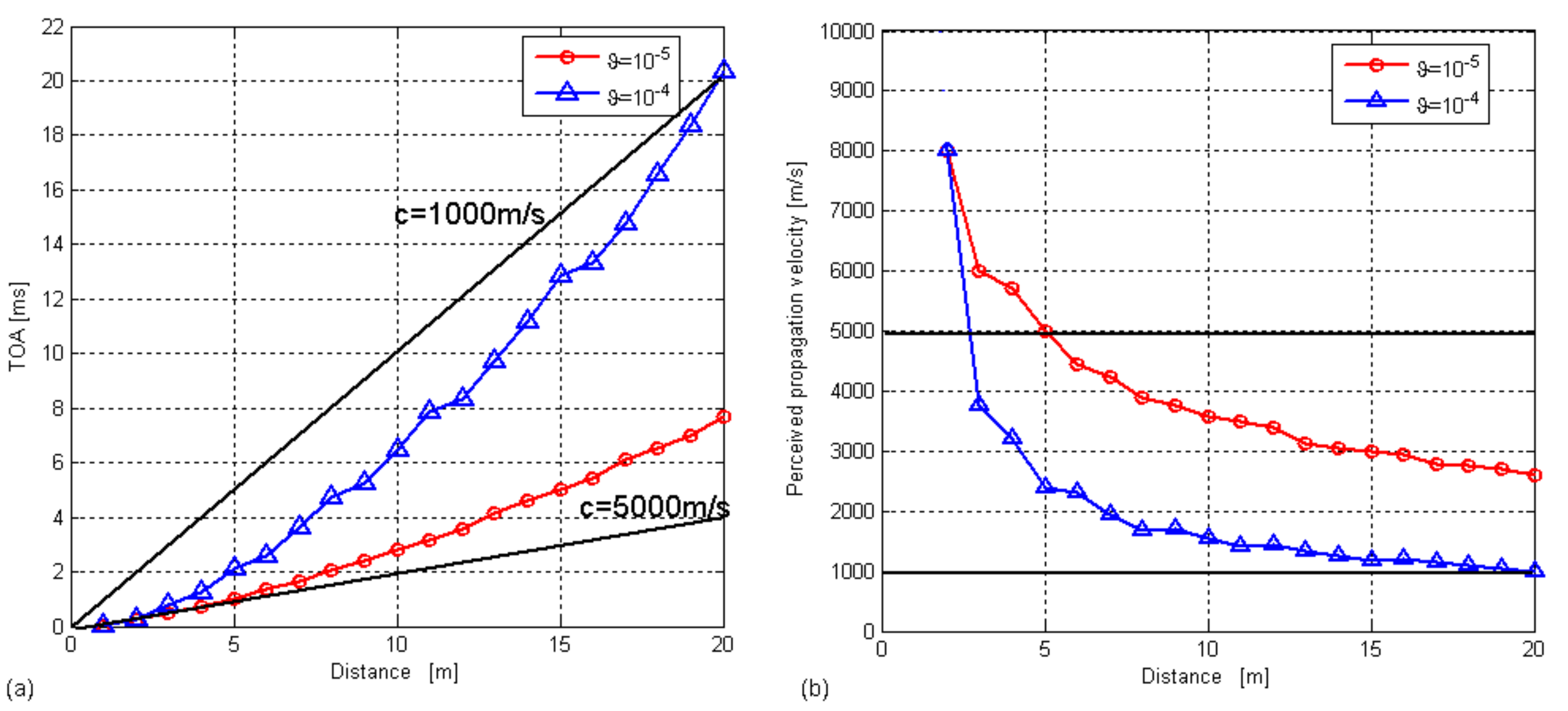}
\centering
\caption{\small{Attenuation and dispersion effects on TOA detection (a) and perceived propagation velocity (b).}}
\label{TOAdcpd1}
\end{figure}

The TOA can be detected when the signal exceeds a certain threshold $\Delta$. Figure \ref{TOAdcpd1} (a) shows the TOA variation as a function of the source-sensor distances for $\vartheta = 10^{-5}$ and $\vartheta = 10^{-4}$. The threshold value is arbitrarily fixed. Figure \ref{TOAdcpd1} (a) also shows the variation of the TOA if the perceived propagation velocity is a constant $c = 1000 \mathrm{m/s}$ and $c = 5000 \mathrm{m/s}$. \\ 

Simulation results show that the TOA ($t_a$) and the source-sensor distance ($d$) are not linearly dependent. The "perceived propagation velocity" $c_p$ is defined by:
\begin{equation}
c_p(d)=\frac{d}{t_a(d)}
\end{equation}
and it is not a constant as a function of the source-sensor distance. However, the order of arrival of the signal at the different sensors is maintained (i.e., the TOA increases when the source-sensor distance increases). Figure \ref{TOAdcpd1} (b) shows that the perceived propagation velocity appears to actually decrease with respect to the propagation distance $d$. So, if two sensors are placed such that sensor 1 is closer to the source $k$, we have:
\begin{equation}
d_{k1}<d_{k2} \quad \Rightarrow \quad c(d_{k1})>c(d_{k2}) 
\end{equation}
where $d_{ki}$ is the distance between the source $k$ and the sensor $i$, and $c(d_{ki})$ is the perceived propagation velocity at the sensors $i$, and then we obtain:
\begin{equation}
t_{k1}(d_{k1})= \frac{d_{k1}}{c(d_{k1})}<t_{k2}(d_{k2})= \frac{d_{k2}}{c(d_{k2})}
\end{equation}
where $t_{ki}$ is the TOA detected at sensor $i$. The TOA detected at the sensor closest to the source is the shortest, i.e. : 
\begin{equation}
\mathrm{If} \quad d_{k1}<d_{k2} \Rightarrow t_{k1}<t_{k2}.
\end{equation} 
Although the approximation of Eq. (\ref{udt}) allows the demonstration of the behavior of the propagation wave in a thin plate, the level of approximation, as well as the nature of the approximation, barely allows the relationship between $d$ and the TOA to be extracted. To approximate the expression of the perceived propagation velocity as a function of the source-sensor distance, we use the stationary-phase approximation method, as in the next paragraph. 
%----------------------------------
\subsection{Perceived propagation velocity - stationary phase approximation}
\label{StatPhaseApprox}
The stationary-phase method allows the approximation of the evaluation of Eq. (\ref{eq7}) in the case of a wave packet that propagates in the medium. This leads to the identification of the central frequency of the wave packet as a function of $d$ and $t$ \cite{DR_1996}. This approximation is more accurate at around the maximum of the signal. We can write $u(d,t)$ as: 
\begin{equation}
\begin{array}{lll}
u(d,t)&= \int U(0,\omega) \mathrm{e}^{-k_I(\omega) d} \mathrm{e}^{j[k_R(\omega) d- \omega t]} \mathrm{d} \omega &= \int F(\omega) \mathrm{e}^{-j\Phi(\omega)} \mathrm{d}\omega\\
\end{array}
\label{form}
\end{equation}
where:
\begin{eqnarray}
F(\omega)&=&{U(0,\omega)} \mathrm{e}^{-k_I(\omega) d} \\
\Phi(\omega)&=&-[k_R(\omega) d- \omega t]
\end{eqnarray}
The stationary phase method consists of expanding $\Phi(\omega)$ in a Taylor series near the point $\omega_0$ of the stationary phase (i.e $\Phi'(\omega_0) = 0$), keeping only the first two nonzero terms:
\begin{equation}
\Phi(\omega)\simeq \Phi(\omega_0)+\frac{1}{2}\Phi^{''}(\omega_0)(\omega-\omega_0)^2
\end{equation}
and approaching $F(\omega)$ by $F(\omega_0)$, the integral in Eq. (\ref{form}) can be approached by:
\begin{eqnarray}
u(d,t) &\simeq & F(\omega_0) \mathrm{e}^{-j\Phi(\omega_0)} \sqrt{\frac{2 \pi}{ j |\Phi^{''}(\omega_0)|}}\\
&\simeq & \frac{U(0,\omega_0)}{\sqrt{d |k_R^{''}(\omega_0)|}} \mathrm{e}^{-k_I(\omega_0) d} \mathrm{e}^{j \left(k_R(\omega_0)d-\omega_0 t +\frac{\pi}{4}\right)}
\end{eqnarray} 
By inserting the stationary phase condition ($\Phi'(\omega_0) = 0$), we get:
\begin{equation}
\omega_0= \left( \frac{d}{2 \sqrt{a} t} \right)^2
\end{equation}
The envelope of the wave can then be calculated as: 
\begin{equation}
\begin{array}{ll}
A(d,t)&=\dfrac{|U(0,\omega_0)|}{\sqrt{d |k_R^{''}(\omega_0)|}} \mathrm{e}^{-k_I(\omega_0) d}= 2 |\alpha | a^{\frac{1}{4}} d^{-\frac{1}{2}} \omega_0^{\frac{5}{4}} \mathrm{e}^{-\dfrac{\vartheta \omega_0^{\frac{3}{2}} d}{4 \sqrt{a}} }\\
&\\
&= \dfrac{|\alpha| a}{2 \sqrt{2}} \dfrac{d^2}{t^{5/2}} \mathrm{e}^{-\dfrac{\vartheta}{32 a^2} \dfrac{d^4}{t^3}}
\end{array}
\label{Amp}
\end{equation}
We want to establish the relation between the perceived propagation velocity and the source-sensor distance. The proposed approach consists of studying the evolution of the maximum of the envelope in time and distance $d$ from the source position. The maximum of the envelope of the signal satisfies $\dfrac{\partial A}{\partial d} = 0$, then:
\begin{equation}
1-\dfrac{\vartheta}{16 a^2 t^3} d^4 = 0
\end{equation} 
and the maximum of the envelope is located at each time at $d$, which is given by: 
\begin{equation}
d=\left[ \dfrac{16 a^2 }{\vartheta} t^3 \right]^{\dfrac{1}{4}}
\end{equation}
In other terms, the TOA of the maximum of the envelop at a distance $d$ is: 
\begin{equation}
t=\left[ \dfrac{\vartheta}{16 a^2} d^4\right]^{\dfrac{1}{3}}.
\end{equation}
The "perceived propagation velocity" can then be calculated as: 
\begin{equation}
\boxed{
c_p=\frac{d}{t}=\frac{d}{\left[ \dfrac{\vartheta}{16 a^2} d^4\right]^{\frac{1}{3}}}= \left[\frac{16 a^2}{\vartheta d}\right]^{\dfrac{1}{3}}.
}
\label{c_p}
\end{equation}
Eq. (\ref{c_p}) shows that the "perceived propagation velocity" is not a constant as a function of the source-sensor distance, as it varies like a constant multiplied by $(1/d)^{1/3}$. Moreover, it shows that the "perceived propagation velocity" decreases when the source-sensor distance increases. These results reinforces those of section \ref{Sim}, which were obtained by a numerical approximation of the integral (\ref{eq7}).\\

\begin{figure}[h!]
\centering
\includegraphics[width=13.5cm]{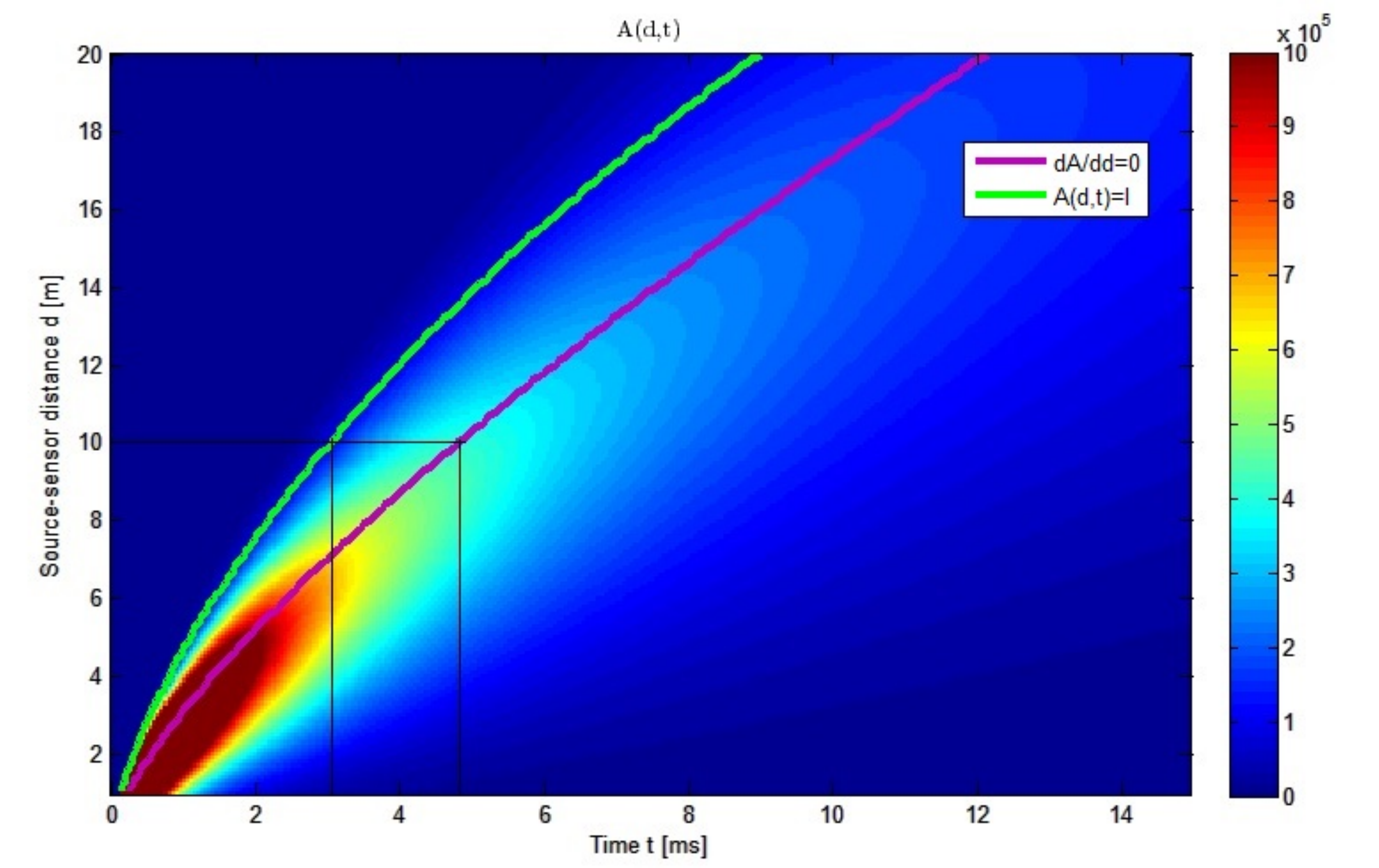} %modifed OM 
\centering
\caption{\small{Envelope of the signal amplitude as function of time and source-sensor distance. Thick line, movement of the maximum of the amplitude in time and distance. Thin line, movement of a point defined by a constant envelope level ($l = 10^5$) in time and distance.}}
\label{A_dt}
\end{figure}

Figure \ref{A_dt} shows the simulated amplitude of the envelope $A(d,t)$ given by Eq. (\ref{Amp}), as a function of time and source-sensor distance ($a = 183$ and $\vartheta = 10^{-5} \mathrm{s}$). It also shows the movement of the maximum of the amplitude (thick line) and the movement of a point defined by a constant envelope level \footnote{Arbitrarily chosen equal to $10\%$ of the maximum of the envelope at $d=5m$.} ($l = 10^5$) (thin line) in time and distance.\\
\begin{figure}[h!]
\centering
\includegraphics[width=13cm]{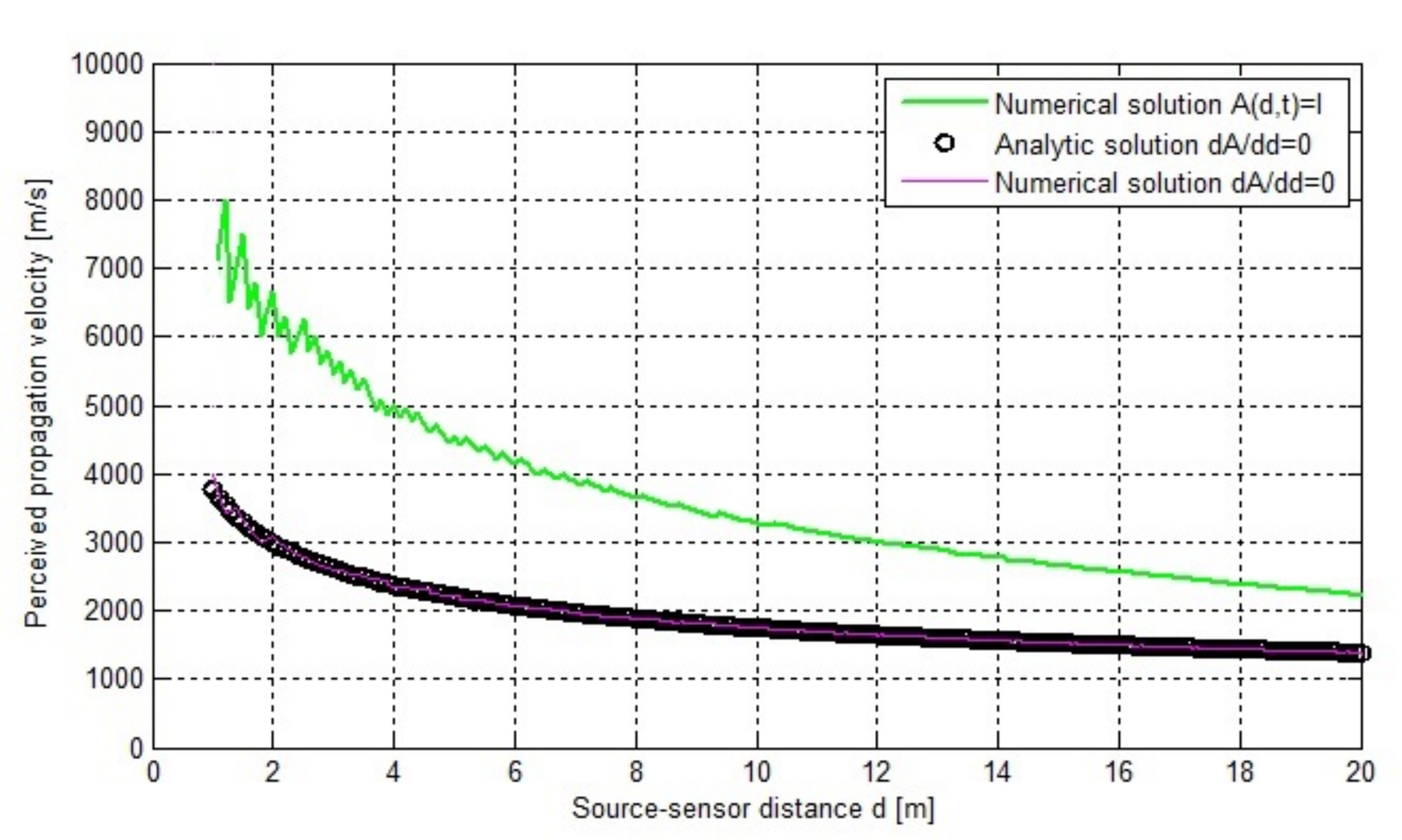}%,height=6.2cm;,    modified OM
\centering
\caption{\small{Perceived propagation velocity. Thin line :  numerical solution of $A(d,t) = 10^5$; thick line : numerical solution of $\dfrac{\partial A}{\partial d} = 0$. Analytic solution $\dfrac{\partial A}{\partial d} = 0$ Eq. (\ref{c_p}), (o).}}
\label{perc}
\end{figure}

Experimentally, for large source-sensor distances, high frequencies are severely damped and the signal is dominated by the low frequency components. Detection of the maximum suffers therefore from high variance. A threshold-based approach can be numerically solved to give the shape of the variation of the perceived propagation velocity as a function of the source-sensor distances, using the envelope of the signal. The analytical solution of the threshold-based approach cannot be easily determined. \\

Figure \ref{perc} shows the "perceived propagation velocity", as determined by the numerical and analytic solutions of the method, solving $\frac{\partial A}{\partial d} = 0$, and by the numerical solution of the method solving $A(d,t) = 10^5$. This concludes that the "perceived propagation velocity" depends on the source-sensor distance.

\subsection{Experimental results}
\label{Sim2}
\begin{figure}[h!]
\centering
\includegraphics[width=10cm,height=6.5cm]{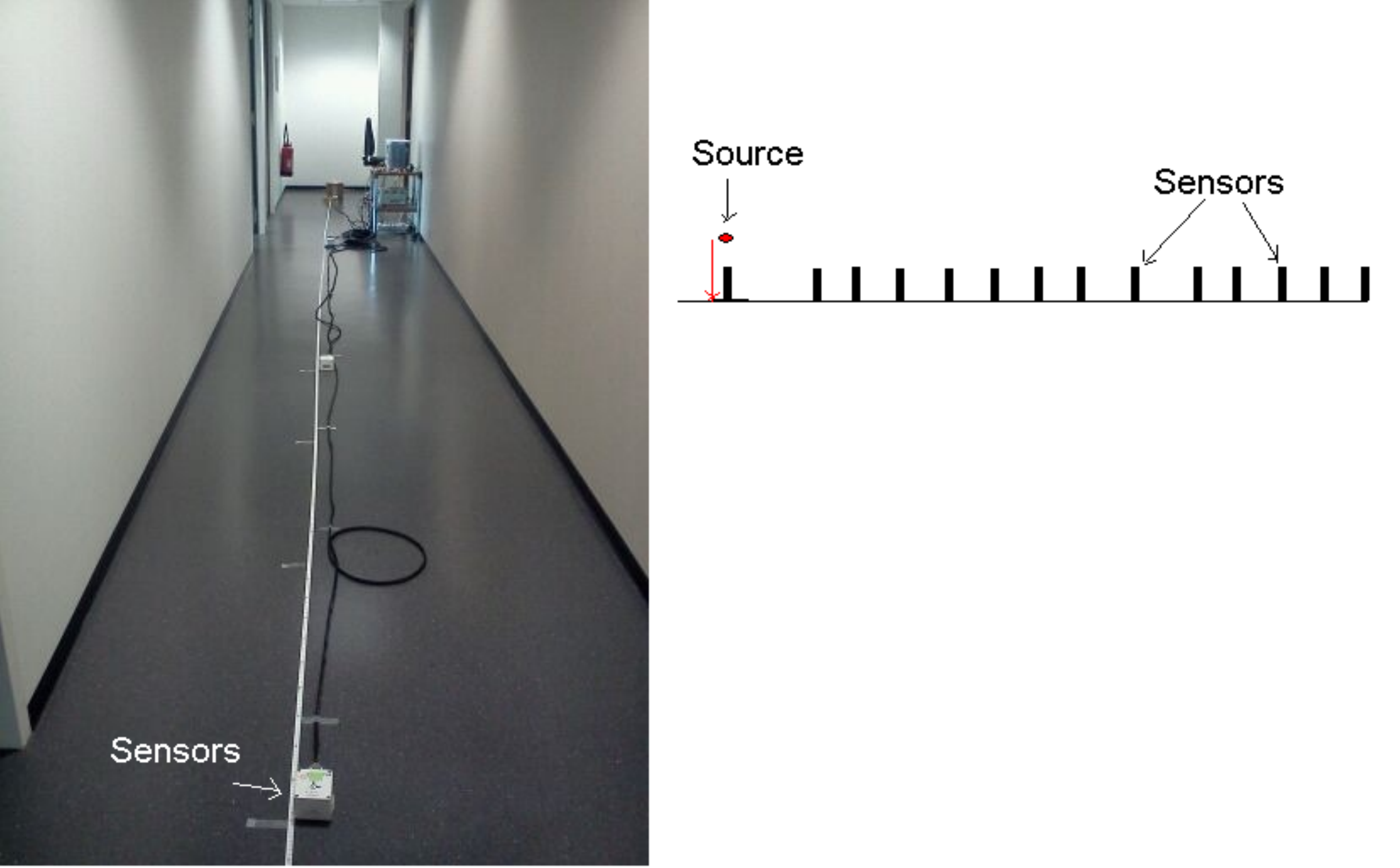}
\centering
\caption{\small{Attenuation and dispersion effects on the perceived propagation velocity: experimental set-up.}}
\label{balle_mul}
\end{figure}

To confirm the theoretical relationship between the perceived propagation velocity and source-sensor distance, an experimental approach was considered. For the experimental results, we use data recorded during indoor tests. The propagation medium considered is a $20\mathrm{cm}$-thick concrete slab covered by linoleum. The sensors (accelerometers) are deployed in a linear array. To characterize the propagation for the medium, we used a reproducible source: a ball was dropped from a height of $1.50 \mathrm{m}$ several times near a reference sensor $g_0$ (Figure \ref{balle_mul}). \\

\begin{figure}[h!]
\centering
\includegraphics[width=11cm]{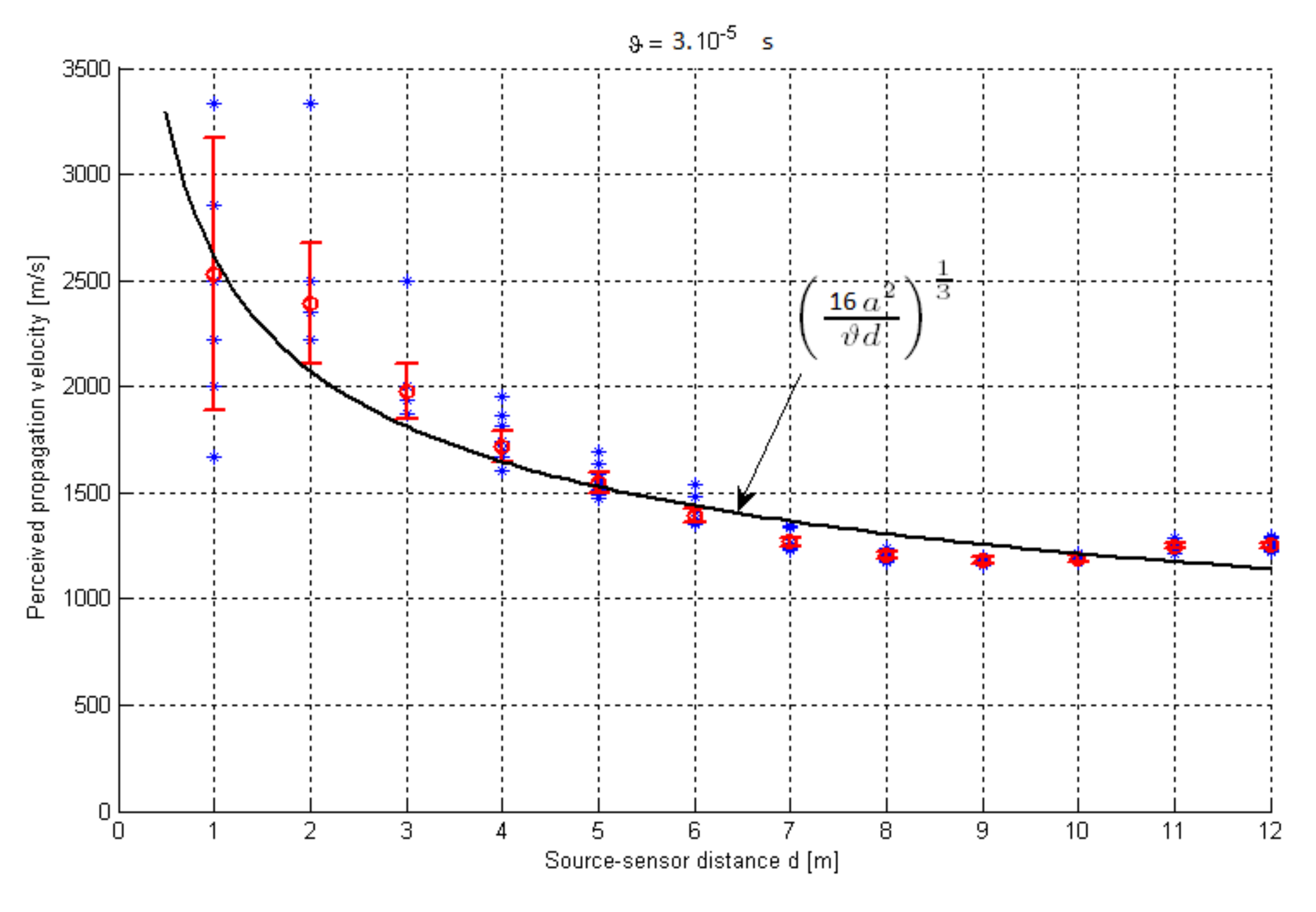}%,height=9cm
\centering
\caption{\small{Attenuation and dispersion effects for the perceived propagation velocity: simulation and experimental results. (*) estimated propagation velocity for one ball drop; (o) mean of all estimated propagation velocities for each distance; ($|$) error for the estimated propagation velocity related to a $0.1$ms ($2$ samples) error on the measured TOA.}}
\label{c_d_2}
\end{figure}

Figure \ref{c_d_2} shows the theoretical "perceived propagation velocity" given by Eq. (\ref{c_p}) for $\vartheta = 3. 10^{-5}\mathrm{ s}$ and the estimated velocity at each sensor by:
\begin{equation}
c_p(d_i)=\frac{d_i}{t_a(d_i)-t_a(d_0)}
\end{equation}
where $d_i$ is the distance between sensor $i$ and sensor $0$, and $t_a(d_i)$ is the estimated TOA at sensor $i$, as determined by the threshold level on the signal. The choice of the threshold level depends on the measured noise level. \\

Figure \ref{c_d_2} shows the similarity between the experimental and theoretical variations of the perceived propagation velocity with distance. It should be noted that the theoretical result corresponds to the movement of the maximum of the amplitude, and that the experimental result is obtained from the movement of the beginning of the signal that exceeds the noise.\\
The experimental results confirm again that the "perceived propagation velocity" decreases when the source-sensor distance increases in a damped and dispersive thin plate. 

Note that eventhough this seems to exhibit a very simple algebraic relation between $d$ and $c$, it involves parameters that experimentally turned out to be highly variable even for close path trajectories. This again forbids to rely on such a model for the localization problem.

%-------------------------
\subsection{Conclusion}
To summarize this section, we have shown that the propagation velocity estimation depends on the damping and dispersion effects and on the source-sensor distance, and we have shown the relation $c_p(d)$. Consequently, source localization techniques based on different range estimations are not applicable. However, we observed that the order of the arrival at the sensors is maintained even in the presence of damping and dispersion. Experimental tests in an indoor environment confirmed these results. However, in some cases, and especially when the floor was not orthotropic due to the presence of beams in its construction, the order might not be maintained.\\ 
Thus a new algorithm based on the sign of time delay promises good localisation results. In the next section we propose a new SO-TDOA algorithm. 
%-----------
%\clearpage
\section{New SO-TDOA algorithm}
\label{Principle_SO-TDOA}
\begin{figure}[h!]
\centering
\includegraphics[width=4cm]{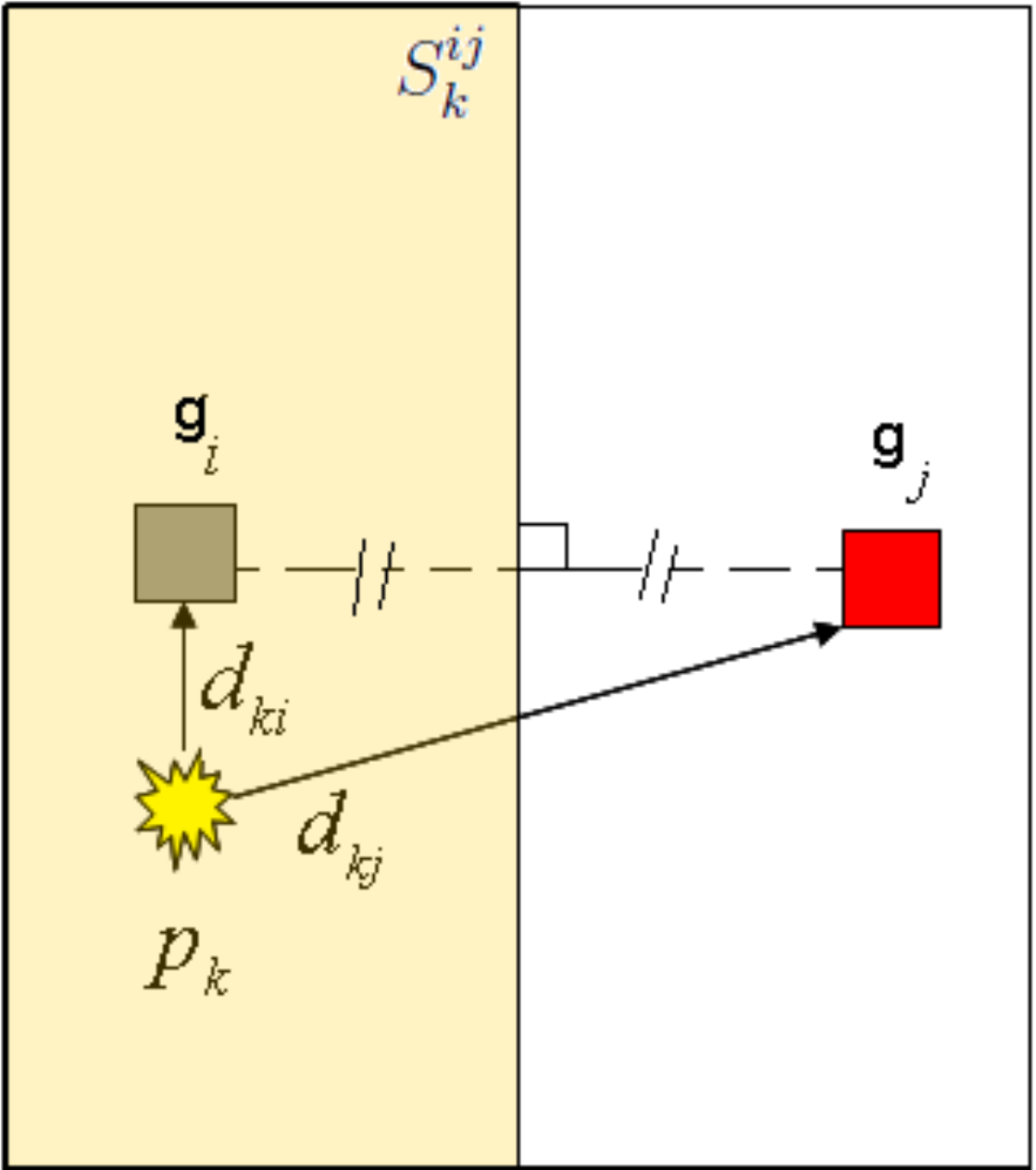}
\centering
\caption{\small{Region separation.}}
\label{deux_cap2}
\end{figure}

Assuming a damped and dispersive floor, the problem of footstep localization in indoor environments cannot be solved using traditional source-localization algorithms based on range estimations, because the perceived propagation velocity depends on the source-sensor distances (section \ref{TDOA}). Furthermore, a received signal strength (RSS) approach cannot lead to acceptable results either, as the coefficient of the parametric model in figure \ref{TOAdcpd1} is highly variable even for close paths on the same slab. Another important issue is the presence of boundaries which induce echoes, modes and interferences. These additional effects preclude any possibility to derive a reliable RSS-based approach. However, the ordering of the arrival time of the signals at different sensors is maintained even in the presence of dissipation and dispersion. In other words, for all of the source positions $\mathbf{p}_k$ and sensor pair $(i,j)$:
\begin{equation}
\mathrm{sgn}\left( t_{ki}-t_{kj} \right)=\mathrm{sgn}\left( \frac{d_{ki}}{c_{ki}} -\frac{d_{kj}}{c_{kj}} \right) =\mathrm{sgn} \left( d_{ki}-d_{kj} \right),
\label{sign2} 
\end{equation}
where $\mathrm{sgn}$ defines the sign operator, and $d_{ki}$ is the distance between the point $ \mathbf{p}_k$ and the sensor $i$, $t_{ki}$ is the TOA of the signal to the sensor $i$, and $c_{ki}$ is the perceived propagation velocity at sensor $i$. Eq. (\ref{sign2}) shows that the sign of the time delay is independent of the elastic wave propagation velocity in the medium. 
Considering a pair of sensors $(i,j)$ and a point $ \mathbf{p}_k$, the set $S^{ij}_k$ of points that satisfy for all $ \mathbf{p'}_k \in S^{ij}_k $, 
\begin{equation}
\mathrm{sgn} \left( d_{ki}-d_{kj} \right)= \mathrm{sgn} \left( d'_{ki}-d'_{kj} \right),
\label{eq_reg}
\end{equation} 
where $d_{ki}$ (resp. $d'_{ki}$) is the distance between sensor $i$, and $ \mathbf{p}_k$ (resp. $ \mathbf{p'}_k$) is the half space delimited by the perpendicular bisectors of the line segment joining the sensors $(i,j)$ and containing $ \mathbf{p}_k$ (see Figure \ref{deux_cap2}).\\

Considering now $N$ sensors placed in a bounded environment $E \subset \Re^2$. Each sensor is located at a known position $\mathbf{g}_i$. The environment is partitioned into $Q$ disjointed regions $R_k$. Each region is limited by the perpendicular bisectors of the line segments joining a pair of sensors. Figure \ref{Exp_reg_vec} illustrates an example of the configuration using $5$ sensors in a square room and in a rectangular room.\\ 
From Eq. (\ref{sign2}) and Eq. (\ref{eq_reg}), we can deduce the following property. For all points $ \mathbf{p}_k$ and $\mathbf{p'}_k$ in a region $R_k$ and for all pairs of sensors $(i,j)$, we can write:
\begin{equation}
\forall \mathbf{p}_k , \mathbf{p'}_k \in R_k, \quad \mathrm{sgn} \left( t_{ki}-t_{kj} \right)= \mathrm{sgn} \left( t'_{ki}-t'_{kj} \right) 
\end{equation}

The SO-TDOA algorithm consists on region localization. In what follows, we propose to characterize each region formed by perpendicular bisector of pairs of sensors. So we will determine the number $Q$ of the obtained regions, the coordinates of their centroid point $\mathbf{p}_k^c$, and their characteristic vector $\mathbf{z}_k$, as defined below. 
\subsection{Region characteristic vector}
\begin{figure}[h!]
\centering
\includegraphics[width = 12cm]{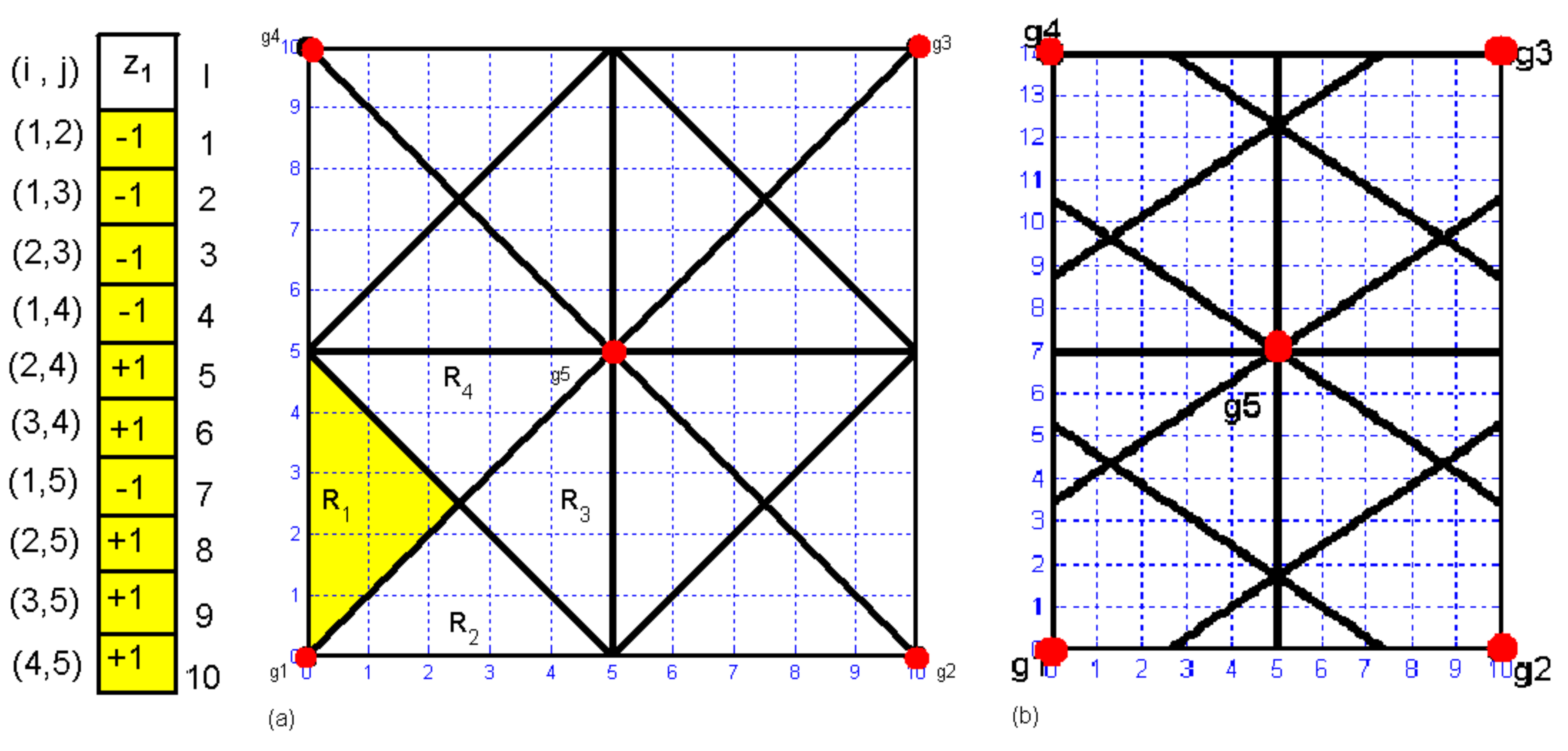}
\caption{\small{Example of region separation: $5$ sensors in a square room (a) and in a rectangular room (b). In the square room, the region $R_1$ is delimited by the perpendicular bisectors of the line segments joining pairs $(1,5)$ and $(2,4)$, and the boundary of the environment. $\mathbf{z}_1$ is the characteristic vector of the region $R_1$ (shaded).}}
\label{Exp_reg_vec}
\end{figure}

Considering all of the sensor pairs $(i,j)$, we can define a characteristic vector $\mathbf{z}_k$ for each region $R_k$ as:
\begin{equation}
\begin{array}{c}
\mathbf{z}_k(l)=\mathrm{sgn} \left( d_{ki}-d_{kj} \right),\quad l=\frac{(j-2)(j-1)}{2}+i 
\label{Zk}
\end{array} 
\end{equation} 
$\forall (i,j) \in \{(1,2),(1,3),(2,3),(1,4),(2,4),\ldots,(N-1,N) \}.$\\
The vector $\mathbf{z}_k$ is formed by $N(N-1)/2$ elements taking values in $\{+1,\-1\}$.\\ 
\textit{Example:} Considering the previous example of configuration, the region $R_1$ can be defined by the vector $z_1$ of $10$ elements, as in Figure \ref{Exp_reg_vec}.\\

\textit{Remarks:} If $N$ sensors are placed in an unbounded plane such that there are no parallel perpendicular bisectors, the number of perpendicular bisectors is equal to $N(N-1)/2$ (i.e., to the number of pairs of sensors). Or considering $o$ nonparallel lines in an unbounded plane, these form $o(o+1)/2+1$ regions. The number of regions formed by $N$ sensors in an infinite space is calculated for $o = N(N-1)/2$ and is obtained as $M_R = (N^4-2N^3+3N^2-2N)/8+1$. $M_R$ is the upper boundary of the number of regions in a bounded plane. Indeed, the number of regions in a bounded plane depends on the number of sensors, their locations, and the room geometry. Therefore, there is no simple expression that gives the number of regions formed in a bounded plane according to a given sensor configuration. For example, the upper bounds of the number of regions formed with $N = 5$ sensors is $M_R = 56$. However the number of regions formed in a square room is $Q = 16$, and in a rectangular room, $Q = 20$ (see Figure \ref{Exp_reg_vec}).\\
The value of $\mathbf{\hat{z}}_k$ is in $\{+1,-1\}^{N(N-1)/2}$ if it is considered that the estimated SO-TDOA might be erroneous for some pairs of sensors. As $2^{N(N-1)/2}\gg Q$, only a few values of $\mathbf{\hat{z}}_k$ actually correspond to one of the acceptable $Q$ regions. For example, in Figure \ref{Exp_reg_vec}, the sensor pairs $(3,4)$ and $(1,2)$ share the same perpendicular bisector, and so the corresponding elements in the characteristic vector must have the same value $+1$ or $-1$. However, under experimental conditions and with the presence of TDOA estimation errors, nonrealistic characteristic vectors can be obtained. Thus, using redundancy in the characteristic vector might lead to improved localization performances. \\
To estimate the set of regions $M_r$ that correspond to a measured characteristic vector $\mathbf{z}_s$, we choose to minimize the Hamming distance between the measured vector and all of the acceptable characteristic vectors $\mathbf{z}_k$. 
\begin{equation}
M_r = \arg \min_{k \in [1..Q]} d_H(\mathbf{z}_s,\mathbf{z}_k) , \quad M_r \subset [1..Q]
\label{ham}
\end{equation}
where $d_H$ is the Hamming distance measuring the number of components that are different in two vectors,
\begin{equation}
d_H(\mathbf{z}_s,\mathbf{z}_k)= \sum_{i=1}^{N(N-1)/2} \left( \mathbf{z}_s(i) \oplus \mathbf{z}_k(i)\right), 
\end{equation}
where $\oplus$ is the exclusive or operator ($a \oplus b = 1$ if $a \neq b$ else $0$, $\forall a,b \in \{+1, -1\})$. The number of regions that minimize the Hamming distances to the measured vector can be $>1$ in some cases where the measured vector $\mathbf{z}_s$ does not correspond to a realistic region according to the sensor configuration considered. This might frequently occur in the presence of TDOA estimation errors. We denote $|M_r|$ as the cardinal numbers of the set $M_r$. Then to have $|M_r|\geq1$ is possible. Below is an example for the region configuration described in Figure \ref{Exp_reg_vec}:
\begin{equation}
\begin{array}{ll}
\mathbf{z}_2&=[-1,-1,-1,-1,-1,+1,-1,+1,+1,+1], \\
\mathbf{z}_1&=[-1,-1,-1,-1,+1,+1,-1,+1,+1,+1],\\
d_H(\mathbf{z}_2,\mathbf{z}_1)&=0 +0 +0 +0+1 +0+0+0+0+0=1.
\end{array}
\end{equation}
Note that two neighboring regions will be "Hamming"-separated by $1$. 

\subsection{Region center coordinates}
\begin{figure}[h!]
\centering
\includegraphics[width = 10cm]{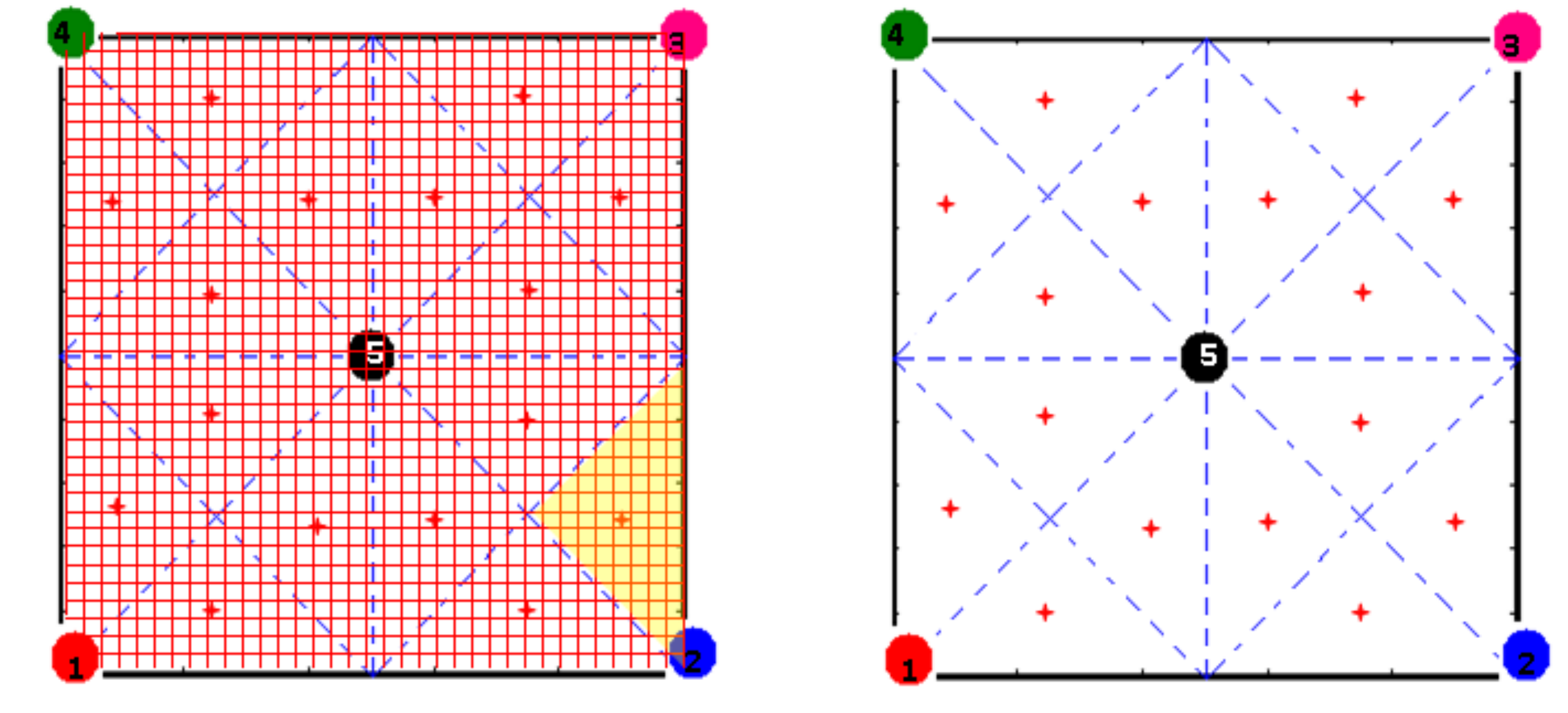}
\centering
\caption{\small{Example: region centroid determination.}}
\label{salle_centre_zone}
\end{figure}

All points located in the same region $R_k$ are characterized by the same vector $\mathbf{z}_k$, as defined by Eq. (\ref{Zk}). All of these points will be associated to their centroid $\mathbf{p}_k^c$. Generally, the geometry of the sensor location (which can be arbitrary) does not allow a simple analytical calculation of the centroid region coordinates to be obtained. We propose to associate each region with its centroid, and to develop a simple computer-based approach to determine its coordinates. This consists of sampling the space with regular points for location $\mathbf{p}_e$; see Figure \ref{salle_centre_zone}.\\ 

For each point $\mathbf{p}_e$, we compute the characteristic vector $\mathbf{z}_e$. Then, all of these points are classified into groups by their characterizing vectors. The number of groups obtained is equal to the number $Q$ of the total regions formed. The centroid coordinates of one region are obtained by averaging the coordinates of all of the points in the same region. This step of the calculation is performed only once, when the sensor configuration is fixed. This information is stored and used later to determine the source position.

\subsection{SO-TDOA localization algorithm}
A human footstep generates a seismic signal that is collected at each sensor in the room. To localize this footstep, the SO-TDOA algorithm is proposed. It consists of the following steps:
\begin{enumerate} 
\item The time of arrival $\hat{t}_{si}$ of the seismic signal at each sensor $i$ is estimated by a simple threshold method. This is determined with respect to a common arbitrary time origin \cite{Maji_1987}.
\item Then the characteristic vector of the source is determined, such that: 
\begin{equation}
\mathbf{z}_s(l)=\mathrm{sgn} \left( \hat{t}_{si}-\hat{t}_{sj} \right), 
\label{zs}
\end{equation}
as arranged in Eq. (\ref{Zk}).
\item The set of regions $M_r \subset [1..Q]$ that minimizes the Hamming distance is estimated: 
\begin{equation}
M_r = \arg \min_{k\in [1..M]} \sum_{i=1}^{N(N-1)/2} \left( \mathbf{z}_s(i) \oplus \mathbf{z}_k(i)\right).
\label{MSE}
\end{equation}
where the cardinal numbers of $M_r$ can be more than one $(|M_r|\geq1)$. 
\item Finally, the source position localization $\mathbf{\hat{p}}_{s}$ is estimated by: 
\begin{equation}
\mathbf{\hat{p}}_{s} = \frac{1}{|M_r|} \sum_{r\in M_r} \mathbf{p}^c_{r} 
\end{equation}
where $\mathbf{p}^c_{r} $ is the centroid of the region $r$. The source position estimate corresponds to the average of the centroids of all of the regions that minimize the Hamming distance to the measured vector. This estimator is a heuristic estimator that will be validated in this study by simulation results. This point will be investigated in more detail in future studies. 
\end{enumerate}
%-----------------------------------------------
\section{Performances studies }
\label{perf}
\begin{figure}[h!]
\centering
\includegraphics[width=14cm]{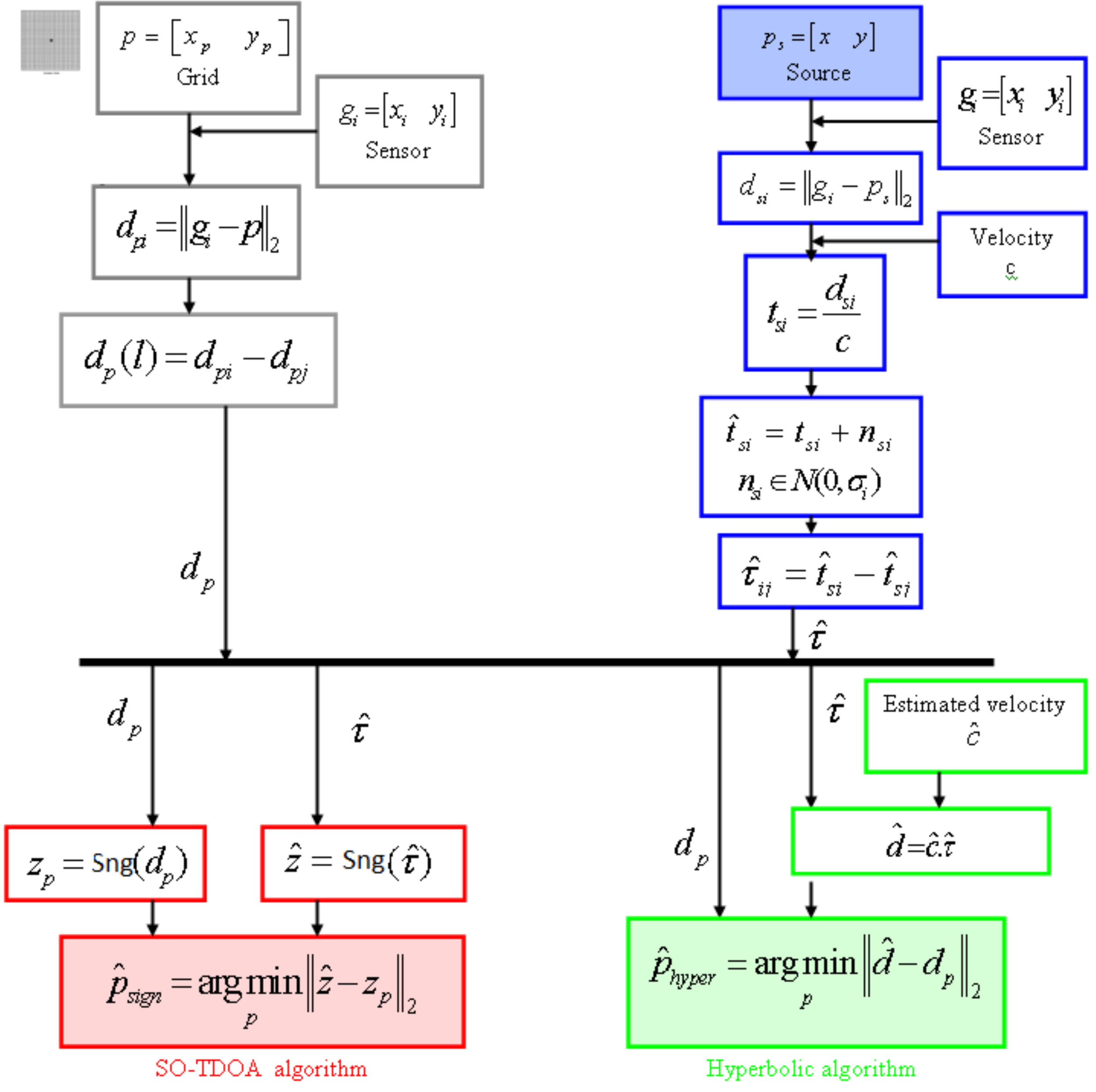}
\caption{\small{Simulation steps.}}
\label{algo}
\end{figure}

In this section, we propose to illustrate the robustness of the proposed SO-TDOA algorithm. We compare it with the classical hyperbolic localization algorithm. When the perceived propagation velocity is assumed to be a constant, the hyperbolic algorithm is one of the best localization algorithms. Theoretically, the perceived propagation velocity depends on the source-sensor distance in a damped and dispersive medium. We indicated that the order of arrival of the signal is maintained, but we have no access to the value of the propagation velocity in each point of a room because it depends on both the attenuation and the dispersion. The values obtained for the estimated propagation velocity might be highly variable, especially in the presence of strong attenuation. The shape of the variation of the perceived propagation velocity versus distances can be as illustrated in Figure \ref{vitess_var2}. \\ 
In this simulation, we study the performances of the two algorithms when the perceived propagation velocity varies, as shown in Figure \ref{vitess_var2}. In concrete, the propagation velocity $c$ can vary from some hundred to some thousand meters per second, depending on the mechanical and physical properties of the medium \cite{Heckl_2010}.\\
For the presented simulations that are based on the hyperbolic algorithm, $\hat{c}$ was set in the range of $500 \mathrm{m/s} \ - \ 2000 \mathrm{m/s}$. These values correspond to reasonable experimentally encountered values. The simulation steps are given in the next paragraph and summarized in Figure \ref{algo}. 

\subsection{Simulation steps}
Figure \ref{algo} illustrates the different steps of the simulation that was conducted to compare the proposed SO-TDOA algorithm with the hyperbolic algorithm \cite{Chan_1994, Zheng_2007}. $N$ sensors are placed in a $L_x \times L_y$ rectangular room, with coordinates $\mathbf{g}_i = [x_i \quad y_i]$, for all $1\leq i \leq N$. \\

\begin{figure}[h!]
\centering
\includegraphics[width = 11cm]{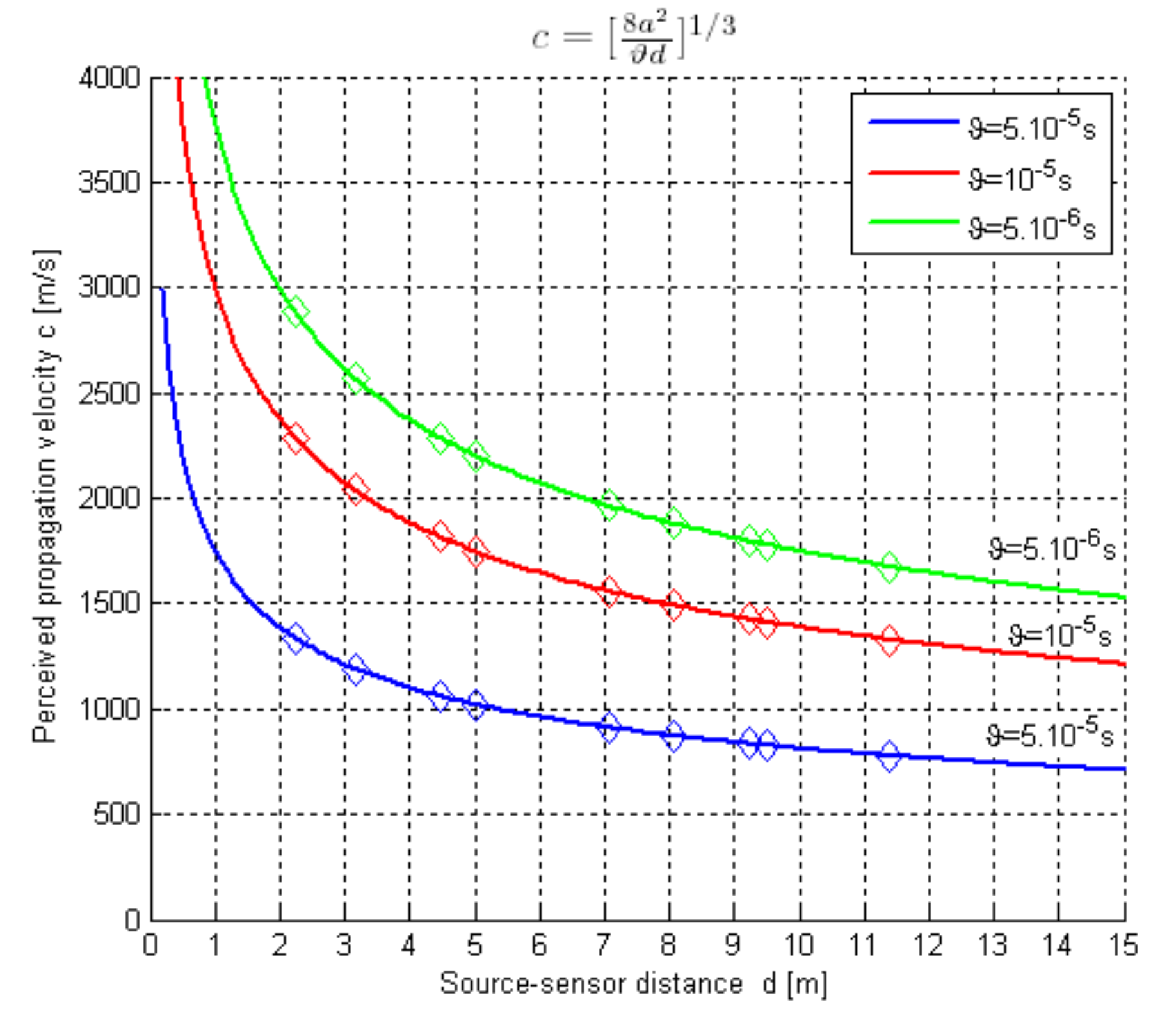}%vitess_var10
\centering
\caption{\small{Shape of the perceived propagation velocity variation versus distance.}}
\label{vitess_var2}
\end{figure}

\underline{Inputs} 
\begin{enumerate}
\item A source position is fixed at $\mathbf{p}_s = [x \quad y]$, such that $0 \leq x \leq L_x$ and $0 \leq y \leq L_y$.
\item The distances from the source to all of the sensors are calculated, as $d_{si} = \|\mathbf{p}_s -\mathbf{g}_i \|_2$, for all $1 \leq i \leq N$. We note $\mathbf{d}_s$, the vector of range differences, such that $\mathbf{d}_s (l) = \left(d_{si}-d_{sj} \right)$ is arranged like in Eq. (\ref{zs}). 
\item Assuming that the perceived elastic-wave propagation velocity is as given by Figure \ref{vitess_var2}, we calculate the arrival times at the sensors as $t_{si} = \frac{d_{si}}{c(d_{si})}$, for all $1 \leq i \leq N$. Under the experimental conditions, we do not have access to the variation in the perceived propagation velocity versus the source-sensor distance. This last closely depends on the properties of the propagation medium. 
\item The times of arrival $t_{si}$ are embedded in an additive zero-mean Gaussian perturbation, with variance $\sigma_i$. We assume that this is white and is independent of the signal or the $t_{si}$. Experimentally, we have obtained a TOA detection error usually in the 
range of $[0 - 1] \mathrm{ms}$ (i.e., for $68\%$ in $[0 - 1] \mathrm{ms}$ this implies $\sigma_t \simeq 0.5 \mathrm{ms}$. An error of $1\mathrm{ ms}$ in the TOA detection implies an error of $0.5 \mathrm{m}$ to $4 \mathrm{m}$ in the source-sensor distance estimation (for propagation velocity $c\in [500-4000] \mathrm{m/s}$).
\item All time delays are determined. Let $\hat{\bm \tau}$ be the vector of $N(N-1)/2$ time delays, such that:
\begin{equation}
\hat{\bm \tau}(l)=\hat{t}_{si}-\hat{t}_{sj}
\label{tau}
\end{equation} 
arranged as in Eq. (\ref{zs}). The vector $\hat{\bm \tau}$ is the input of the localization algorithms. 
\end{enumerate}

\underline{Algorithms:}\\

Both the hyperbolic and SO-TDOA algorithms take $\hat{\bm \tau}$ as their input. The hyperbolic algorithm requires multiple operations to invert the problem. We generate a grid of points that are uniformly distributed in the room, and we search for the point that minimizes the criteria corresponding to the hyperbolic algorithm and the point that minimizes the criteria of the SO-TDOA algorithm: 
\begin{enumerate}
\item [6)] We generate a regular grid of points $\mathbf{p} = [x_p \quad y_p]$ uniformly distributed in the room. 
\item [7)]For each point of the grid, we calculate the range differences vector $\mathbf{d}_p$, such that $\mathbf{d}_p (l) = \left(d_{pi}-d_{pj} \right)$ arranged as in Eq. (\ref{zs}) and $d_{pi} = \|\mathbf{p}-\mathbf{g}_i \|_2$, for all $1 \leq i \leq N$. 
\item [8)]We estimate the source position by the new algorithm based on sign of time delay estimation $\mathbf{\hat{p}}_{s_{|\mathrm{sign}}}$ and by the hyperbolic algorithm $\mathbf{\hat{p}}_{s_{|\mathrm{hyper}}}$.\\
New SO-TDOA algorithm:\\
8.1. For all of the points $\mathbf{p}$, we calculate the vector $\mathbf{z}_p$, such that
\begin{equation}
\mathbf{z}_p (l)=\mathrm{sgn} \left( \mathbf{d}_p(l) \right), \quad \forall l \in [1 \cdots N(N-1)/2].
\end{equation}
8.2. The source position estimated is then given by
\begin{equation}
\mathbf{\hat{p}}_{s_{|\mathrm{sign}}}= \arg \min_{\mathbf{p}} \sum_{i=1}^{N(N-1)/2} \left( \mathbf{\hat{z}}(i) \oplus \mathbf{z}_p(i)\right), 
\end{equation}
where $\mathbf{\hat{z}} = \mathrm{sgn}( \hat{\bm \tau})$.\\
Hyperbolic algorithm:\\
8.1. We calculate $\hat{\mathbf{d}} = \hat{c} \hat{\bm \tau}$, where $\hat{c}$ is a mean propagation velocity that is assumed\footnote{ This latter may be in some simple cases estimated beforehand, assuming a reproducible source and a known location. This is actually far for being realistic in our context. }to be estimated beforehand from a known source location and estimated time delay, see e.g. \cite{Richman_2001}.\\
8.2. The source position estimated is then given by 
\begin{equation}
\mathbf{\hat{p}}_{s_{|\mathrm{hyper}}}= \arg \min_{\mathbf{p}} \|\mathbf{\hat{d}}-\mathbf{d}_p\|_2.
\end{equation}
\end{enumerate}
\subsection{Simulation results}
\begin{figure}[h!]
\centering
\includegraphics[width=10cm]{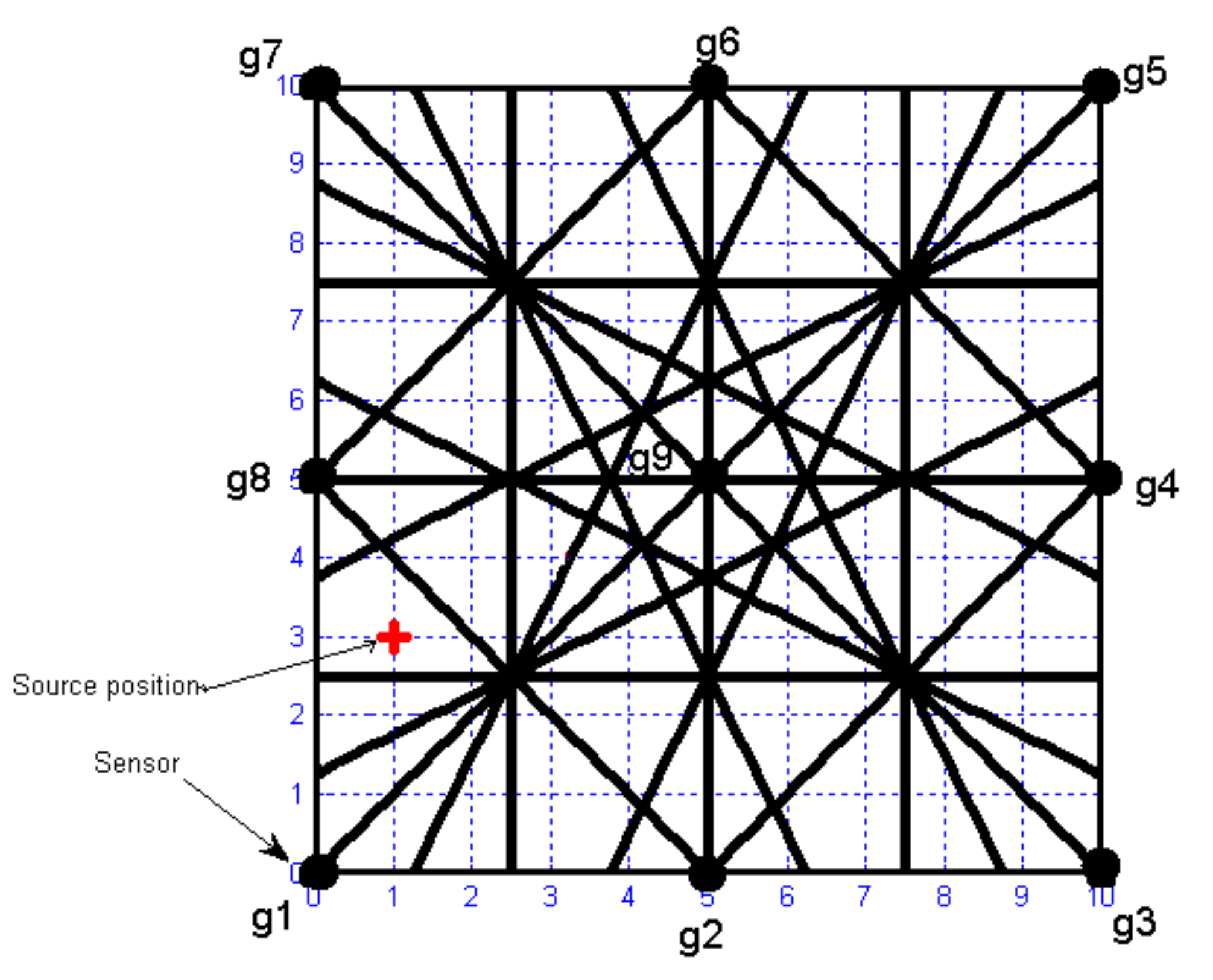}
\caption{\small{Configuration study: Nine sensors positioned in a $10 m\times 10 m$ room. Source positions ($*$).}}
\label{conf}
\end{figure}

We consider $9$ sensors placed in a room $10 \mathrm{m}\times 10 \mathrm{m}$. A source positions is chosen arbitrarily for this study $\mathbf{p}_s = [1 \ 3]\mathrm{m}$, as in Figure \ref{conf}. The grid of points needed for the localization algorithms is generated using $25\times 25$ regular points. The performance index that we use is the root mean squared error (RMSE) between the estimated $\mathbf{\hat{p}}$ and the actual position $\mathbf{p}_s$, as Eq. (\ref{RMSE}). \\
 
\begin{figure}[h!]
\centering
\includegraphics[width=7cm]{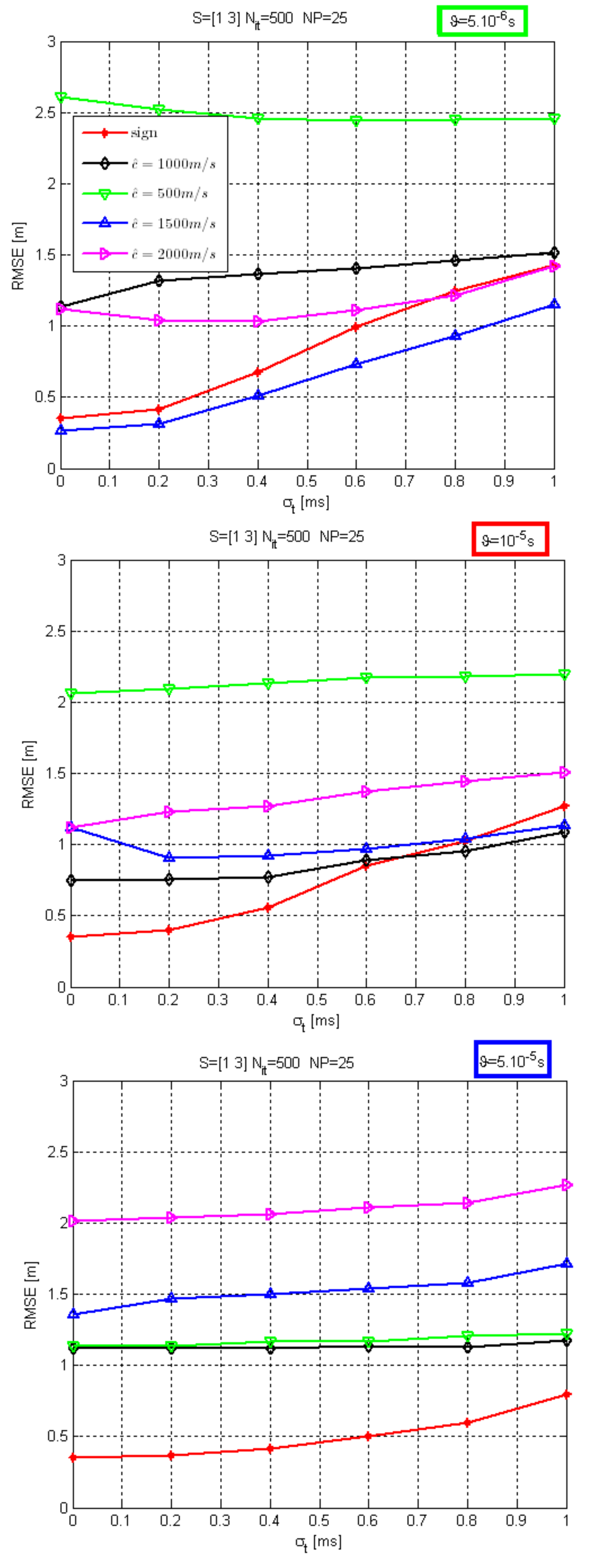}
\caption{\small{Performance study at the source position $\mathbf{p}_s = [1 \ 3]\mathrm{m}$. Perceived propagation velocity varys versus distance, as in Figure \ref{vitess_var2}.}}
\label{perf_1_3_var} 
\end{figure}

Figure \ref{perf_1_3_var} shows the RMSE of the estimated position as a function of $\sigma_t$ for three different shapes of perceived propagation velocity variation (Figure \ref{vitess_var2}) at the same source position. $\sigma_t$ is the standard deviation of the noise simulating TOA estimation errors. Results are obtained by averaging over $M_c = 500$ Monte Carlo runs for all of the investigated scenarios, as for Eq. (\ref{RMSE}). \\
\begin{equation}
\mathrm{RMSE}(\sigma_t)=\sqrt{\frac{1}{M_c}\sum_{i=1}^{M_c} \| \mathbf{p}_s - \hat{ \mathbf{p}}_i(\sigma_t) \|^2 }
\label{RMSE}
\end{equation}

Figure \ref{perf_1_3_var} shows that the proposed localization algorithm SO-TDOA can achieve good localization results ($\mathrm{RMSE} <1.5\mathrm{m}$) even at high TOA estimator error ($\sigma_t = 1 \mathrm{ms}$) in a room of $10 \mathrm{m} \times 10 \mathrm{m}$ without the need for propagation velocity estimation.\\

The hyperbolic algorithm performance depends on the velocity estimation. For different shapes of variation of the perceived propagation velocity, the performances of the hyperbolic  algorithm are changing. We observe that the SO-TDOA algorithm is more robust versus a changing velocity . \\

Finally, it is important to note that the proposed SO-TDOA algorithm is more rapid and has a lower calculation cost compared to the hyperbolic localization algorithm. 
%------------------------------------
\clearpage
\section{Experiment results}
\label{exp}
\subsection{Test set-up}
To validate and assess the performances of the newly developed algorithm, we used data recorded during a series of indoor tests. The soil is a $20 \mathrm{cm}$ concrete slab covered by a tiled floor.
\begin{figure}[h!]
\begin{minipage}[h]{0.5\linewidth}
\centering
\includegraphics[width=7.2cm]{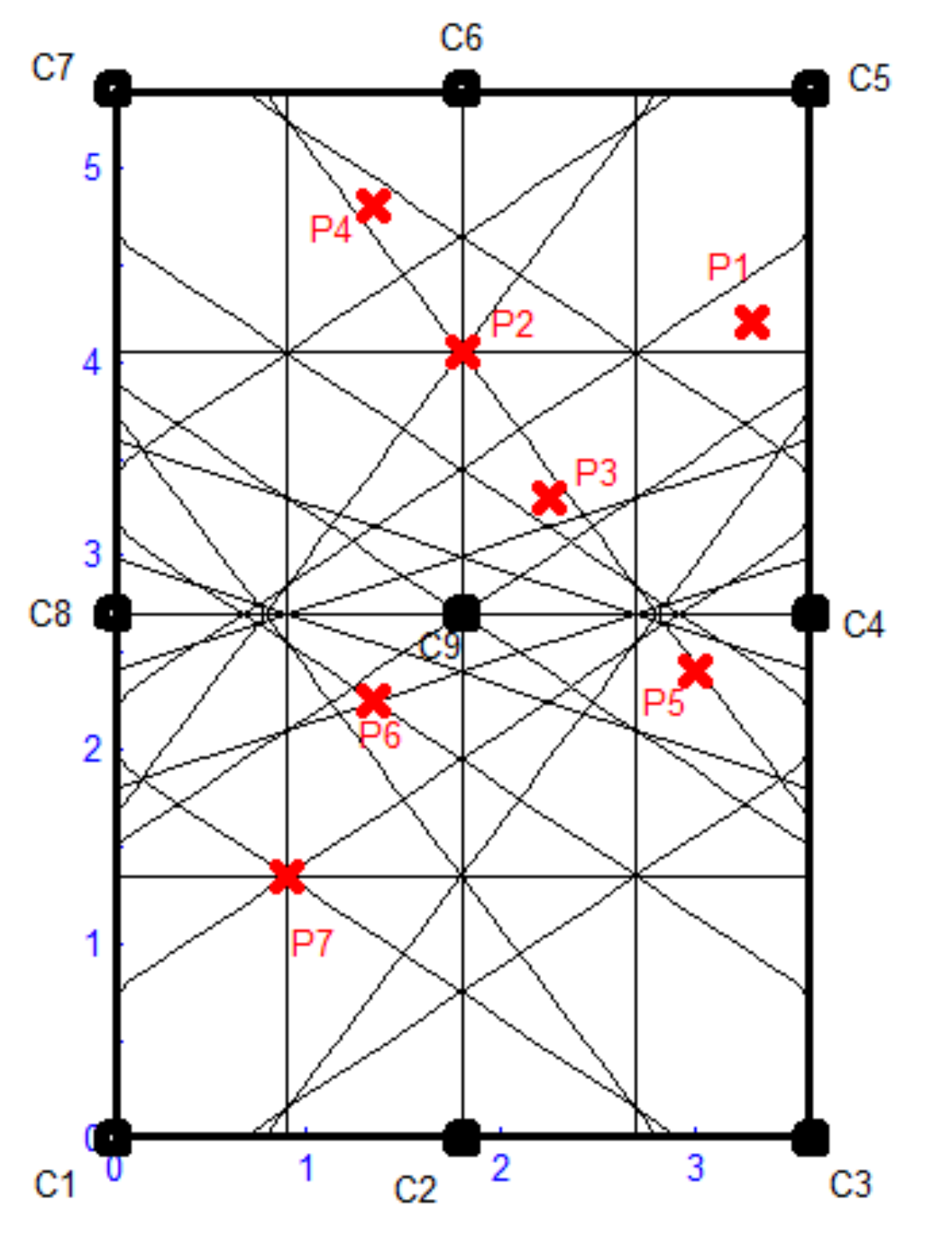}
\end{minipage}
\begin{minipage}[h]{0.5\linewidth}
\centering
\includegraphics[width=8cm]{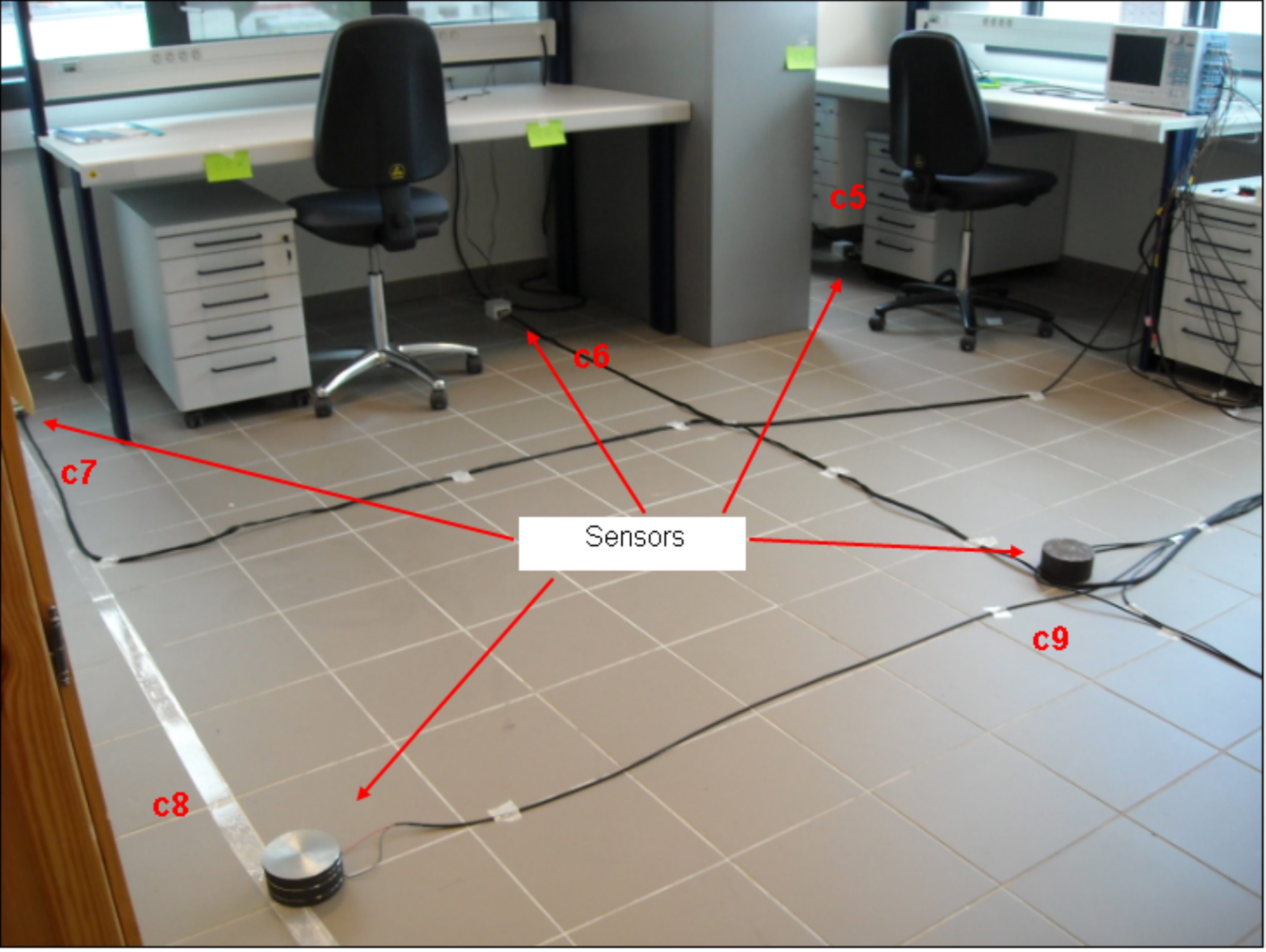}\\
\vspace{0.5cm}
\begin{tabular}{|c|c|c|c|c|c|c|c|}
\hline Footstep & P1 &P2 &P3 &P4 &P5 &P6 &P7\\ 
\hline Position & $3.3$ &$1.8$&$2.25$&$1.35$ &$3$ &$1.35$ &$0.9$\\ 
$[\mathrm{m}]$ & $4.2 $ &$4.05 $&$3.3$ &$4.8$ &$2.4$ &$2.25$ &$1.35$\\ 
\hline 
\end{tabular}
\end{minipage}
\caption{\small{Experimental environment: nine sensors in a rectangular room.}}
\label{exp_loc_zone} 
\end{figure}
As shown in Figure \ref{exp_loc_zone}, nine seismic sensors where placed in a rectangular array on a room $3.6 \mathrm{m} \times 5.4 \mathrm{m}$. Two types of sensors (accelerometers) were used: six piezo-electric ceramics fixed on the floor, with a weight of $5 \mathrm{kg}$, and three Colibry SF3000L fixed with double-faced tape \cite{SF3000L}. The seismic data was acquired, digitized, and relayed to a mobile data-recording station (YOKOGAWA \cite{yokogawa}). The seismic data were sampled at $20 \mathrm{kHz}$. Seven footsteps were monitored in the location giving in Figure \ref{exp_loc_zone}. 
\subsection{Example of experimental signals}
\begin{figure}[h!]
\centering
\includegraphics[height=7cm,width=12cm]{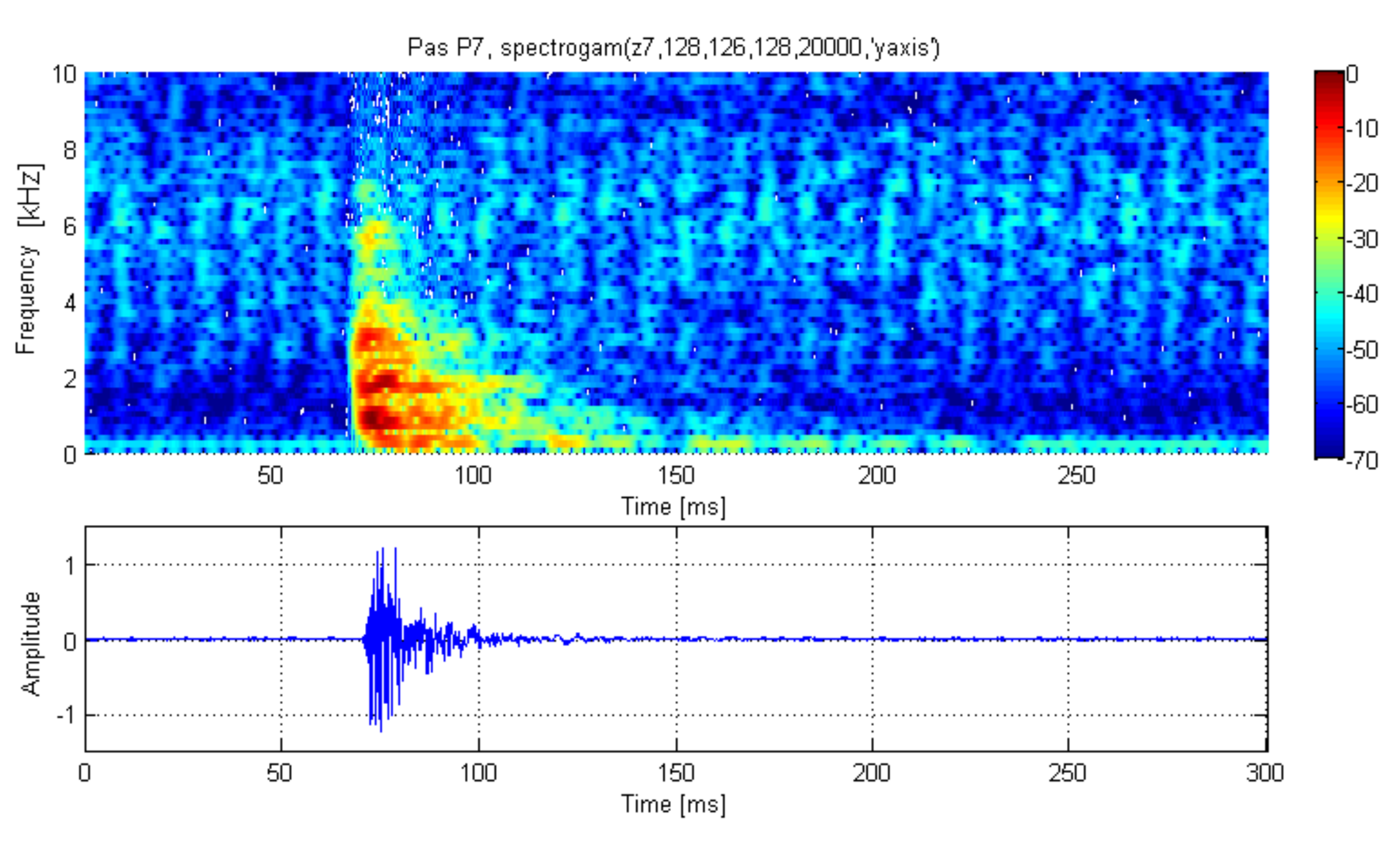}%,height=9cm
\centering
\caption{\small{Signal and short-term Fourier transform of footstep $P7$, received at sensor $C7$. }}
\label{P7}
\end{figure}

\begin{figure}[h!]
\centering
\includegraphics[width=9cm]{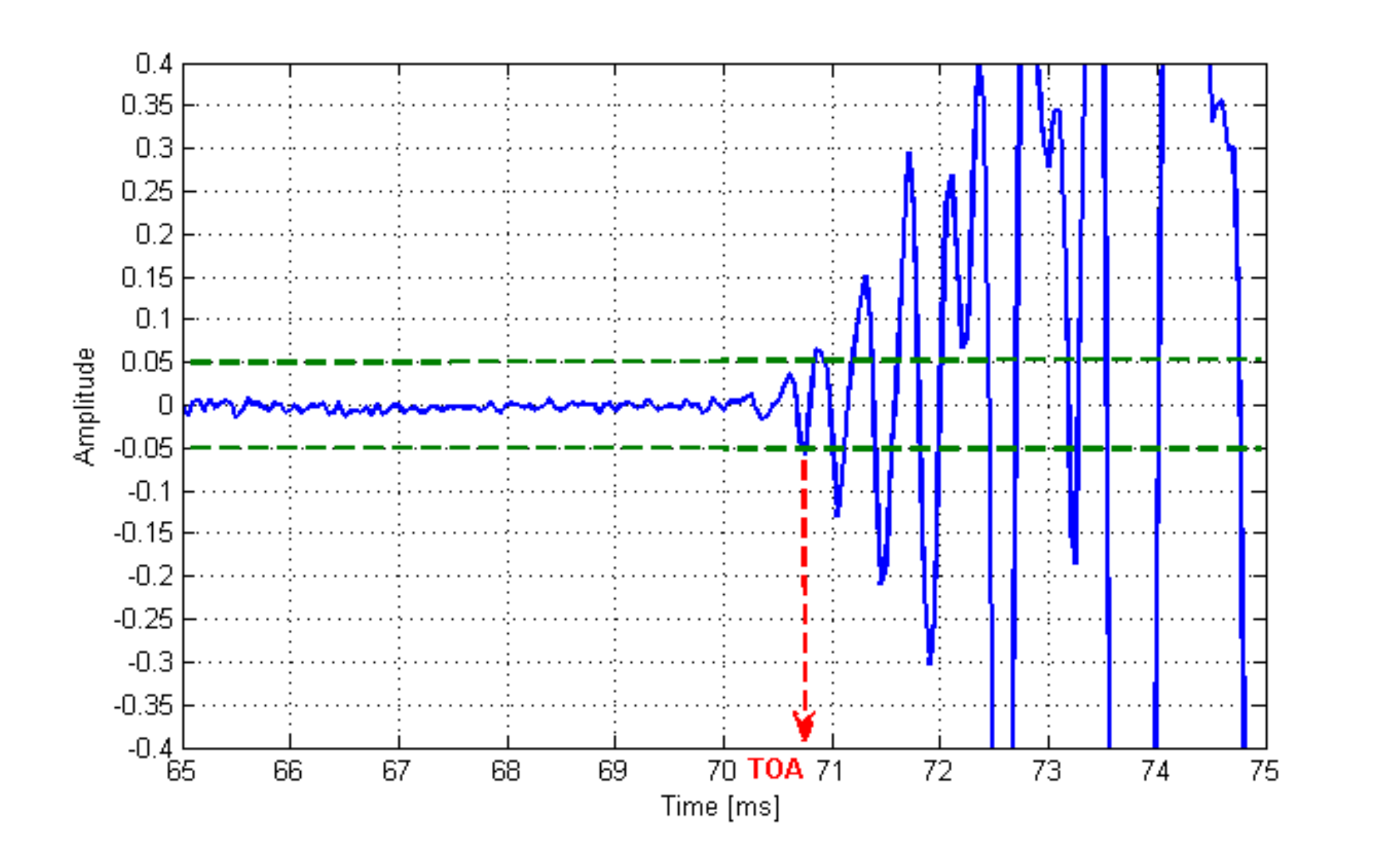}%,height=6cm
\centering
\caption{\small{Example of time-of-arrival detection. A zoom of the signal of footstep $P7$ received at sensor $C7$. }}
\label{P7_zoom}
\end{figure}

An example of a footstep signal and its time frequency representation are given in Figure \ref{P7}. The signal considered corresponds to footstep $P7$ measured at sensor $C7$ (source-sensor distance, $4.15 \mathrm{m}$). The parameters of the short-term Fourier transform are for a Hamming window of length $128$ ($6.4 \mathrm{ms}$), an overlapping segment length of $126$, a fast Fourier transform length of $128$, and a sampling frequency of $20 \mathrm{kHz}$.
We observe that the time frequency representation of the experimental signal in Figure \ref{P7} shows similarity to those of the damping and dispersive medium response [\ref{TFA}]. This implies that the assumption (for a damping and dispersive floor) that is considered in this study conforms to the experimental results. Figure \ref{P7_zoom} shows an example of TOA detection for a seismic footstep signal. 
\subsection{Results}
The position estimation errors of the experiment source are given in Table \ref{Erreur_loc}. The source position estimation error is around some tens of centimeters in a room of $3.6 \mathrm{m}$ by $5.4 \mathrm{m}$.
\begin{table}[h!]
\begin{center}
\begin{tabular}{|c|c|c|c|c|c|c|c|}
\hline Footstep & P1 & P2& P3&P4&P5&P6&P7 \\ 
\hline Estimation error [m]& $ 0.68$ & $ 0.42$& $ 0.54$& $ 0.33$& $ 0.36$& $ 0.29$& $ 0.57$\\ 
\hline 
\end{tabular}
\caption{\small{Experiment source position estimation error using the SO-TDOA algorithm}}
\label{Erreur_loc}
\end{center}
\end{table}
\section{Conclusions}
\label{conc}
In this study, we have proposed a new footstep localization algorithm based on the SO-TDOA. The SO-TDOA algorithm is easily implemented, and it does not need elastic wave velocity estimation. Indeed, we first showed that the elastic wave propagation velocity varies importantly with the source position in an indoor environment, where the floor can be defined as a thin damped and dispersive plate. Using techniques based on range estimation, like a hyperbolic algorithm, it is not sufficient to estimate the footstep position using seismic sensors. The proposed SO-TDOA algorithm provides good simulated and experimental results (a position estimation error of only some tens of centimeters). In future studies, we will adapt SO-TDOA to the dynamic localization of a person in an indoor environment. 
%----------------
\begin{appendix}
\section{Appendix : Time-frequency analysis }
\label{TFA}
\begin{figure}[h!]
\centering
\includegraphics[width=14cm]{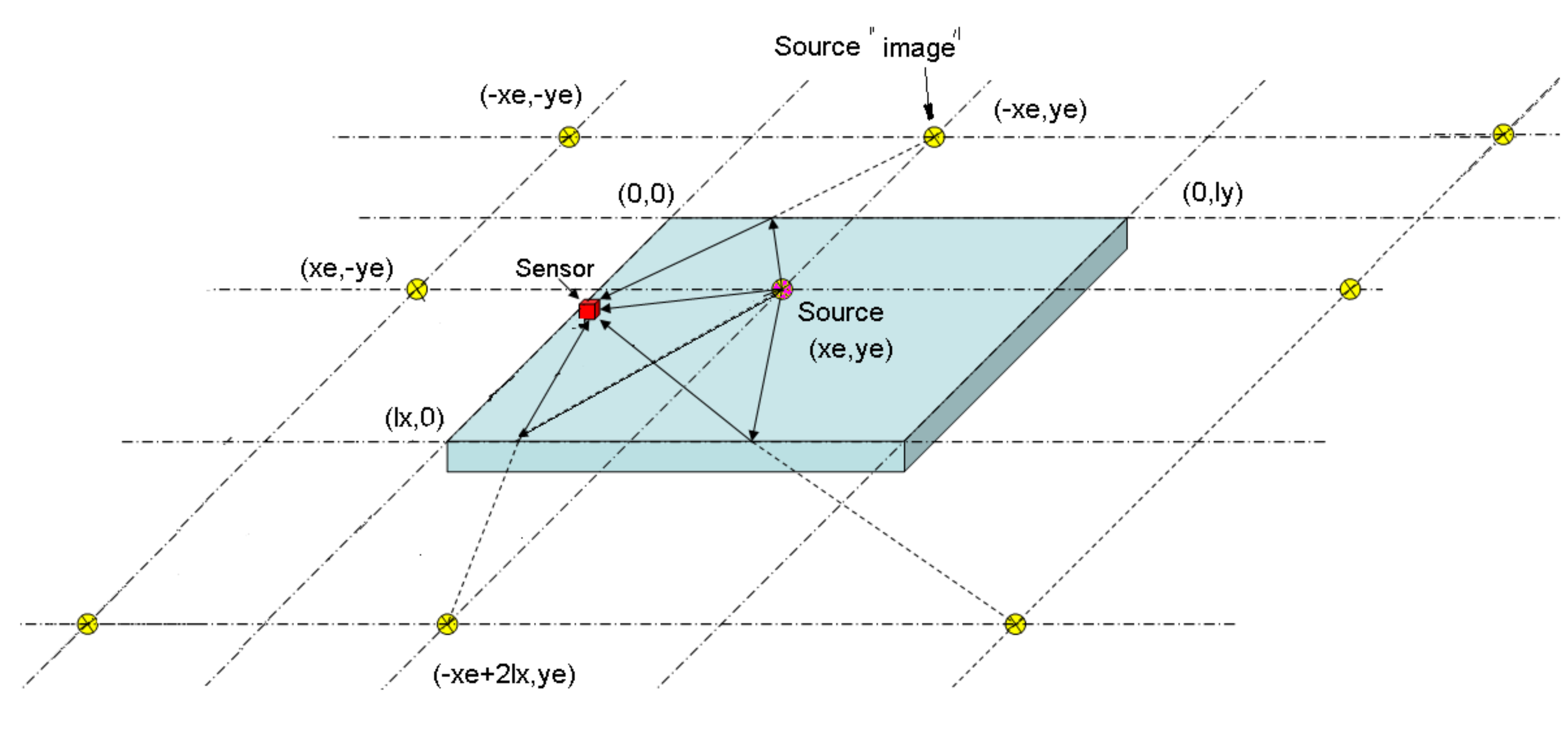}
\centering
\caption{\small{Examples of source "images".}}
\label{SI}
\end{figure}

In this section, we will simulate the time-frequency response of a signal that is propagated in a dissipative and dispersive media. We consider a slab of thickness $0.2 \mathrm{m}$, Young's modulus $E = 24. 10^{-9}\mathrm{N/m^2}$, and mass density $\rho = 2500 \mathrm{kg/m^3}$ \cite{Heckl_2010}. The slab is rectangular, and of width $l_x$ and length $l_y$ in the direction of $x$ and $y$, respectively. A sensor is placed at $[x_r \ \ y_r]$ and a source is placed at $[x_e \ \ y_e ]\mathrm{m}$. Assuming that the edges of the slab are sealed. So reflection induces a change in direction of the displacement. The received signal is the results of successive reflections at the edges. We can represent these reflections from source "images", as for Figure \ref{SI}.
The x and y positions of the source images are given by the two sets:\\
For $x$: $[x_e+2pl_x \quad \quad -x_e+2pl_x] \quad \{p \in [-\infty \ +\infty ], \ \mathrm{integer}\}$ \\ 
For $y$: $[y_e+2ql_y \quad \quad -y_e+2ql_y ]\quad \{ q \in [-\infty \ +\infty ] ,\ \mathrm{integer}\}$ \\ 
The first series underwent an even number of bounces. The signals from these sources should be multiplied by $R_{pq} = 1$. The second should be multiplied by $R_{pq} = -1$.\\
We calculate for $p \leq P_{\max}$ and for $q \leq Q_{\max}$ the distance between the source image $(p,q)$ and the sensor, denoted by $d_{pq}$. We deduce the signal received at the sensor using: 
\begin{equation}
u(t)= \frac{1}{\sqrt{d_{pq}}} \sum_{p,q}^{P_{\max},Q_{\max}} R_{pq} \ u(d_{pq},t)
\end{equation}
where $u(d_{pq},t)$ is given for $d = d_{pq}$ (according to Eq. (\ref{udt})) 
\begin{equation}
u(d,t) \simeq \frac{\omega_m}{n} \sum_{i=0}^{n-1} \mathrm{e}^{-k_I\left( \frac{i\omega_{m}}{n} \right) d} \mathrm{cos} \left( k_R \left( \frac{i\omega_{m}}{n} \right) d- \frac{i\omega_{m}}{n} t \right) 
\end{equation}
where $\omega_{m} = 2 \pi f_{\max}$. $f_{\max}$ is fixed as $10 \mathrm{kHz}$ and $n = 2024$ for this simulation.\\

\begin{figure}[h!]
\centering
\includegraphics[width=14cm]{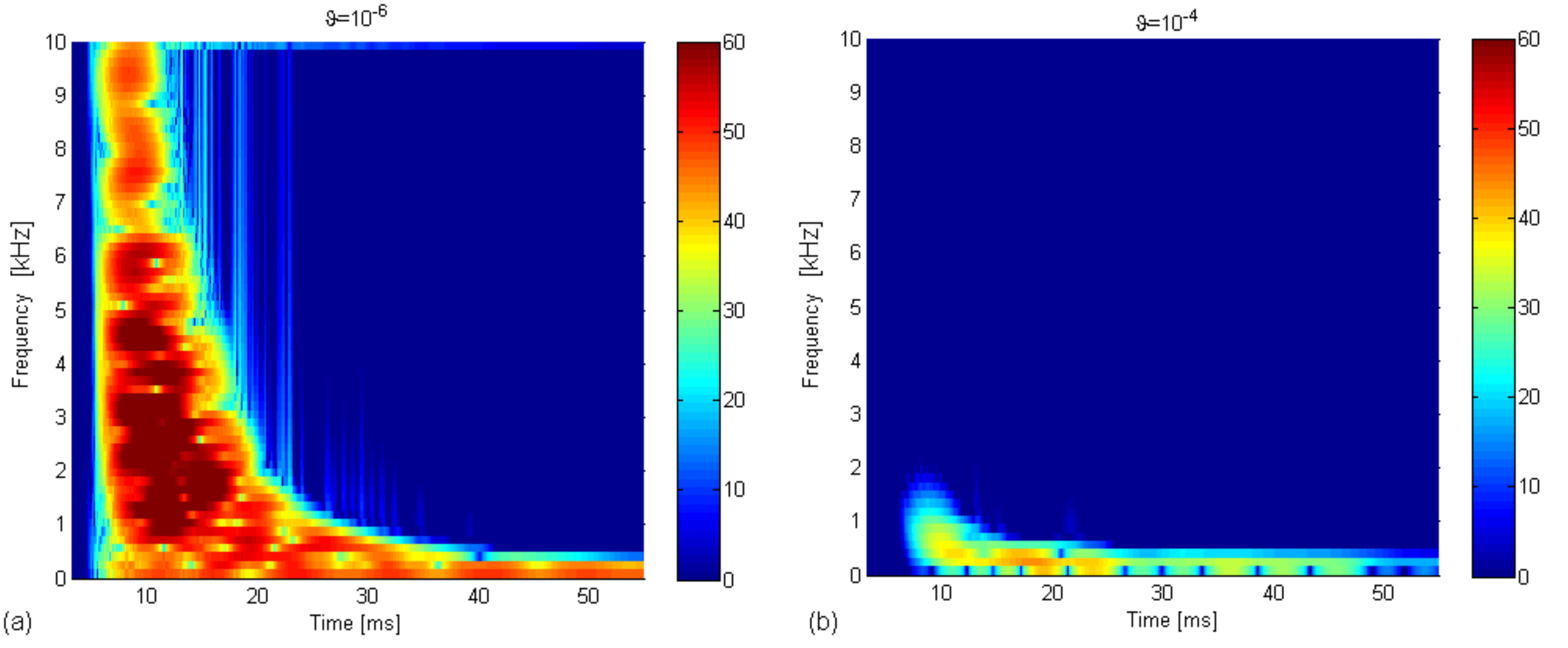}
\centering
\caption{\small{Simulated spectrogram :(a). $\vartheta = 10^{-6}\mathrm{s}$; (b). $\vartheta = 10^{-4}\mathrm{s}$.}}
\label{spec_demo2}
\end{figure}

\begin{figure}[h!]
\centering
\includegraphics[width=10cm]{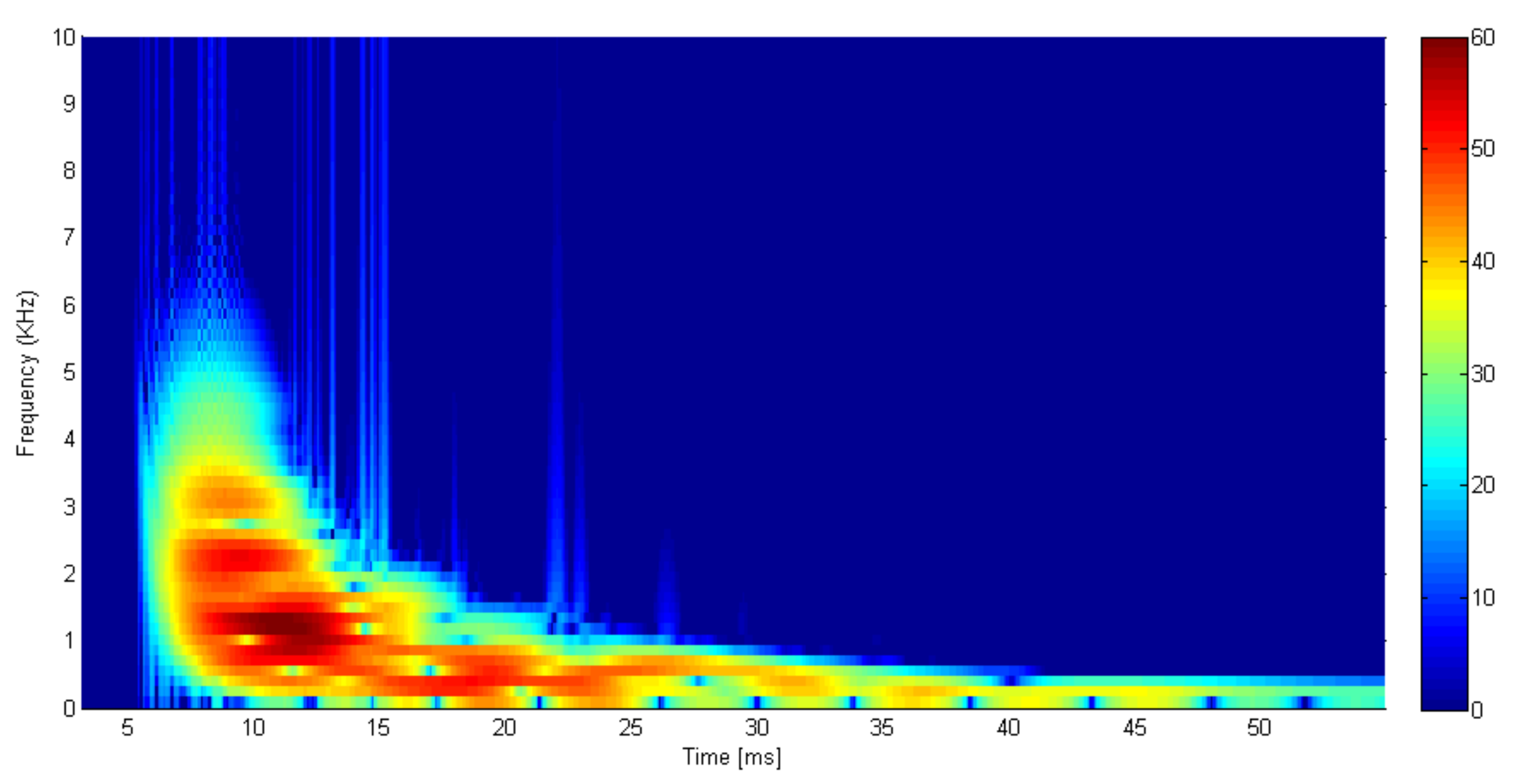}
\centering
\caption{\small{Simulated spectrogram ($\vartheta = 10^{-5}\mathrm{s}$).}}
\label{spec_demo}
\end{figure}

\begin{figure}[h!]
\centering
\includegraphics[height=3.7cm, width=10cm]{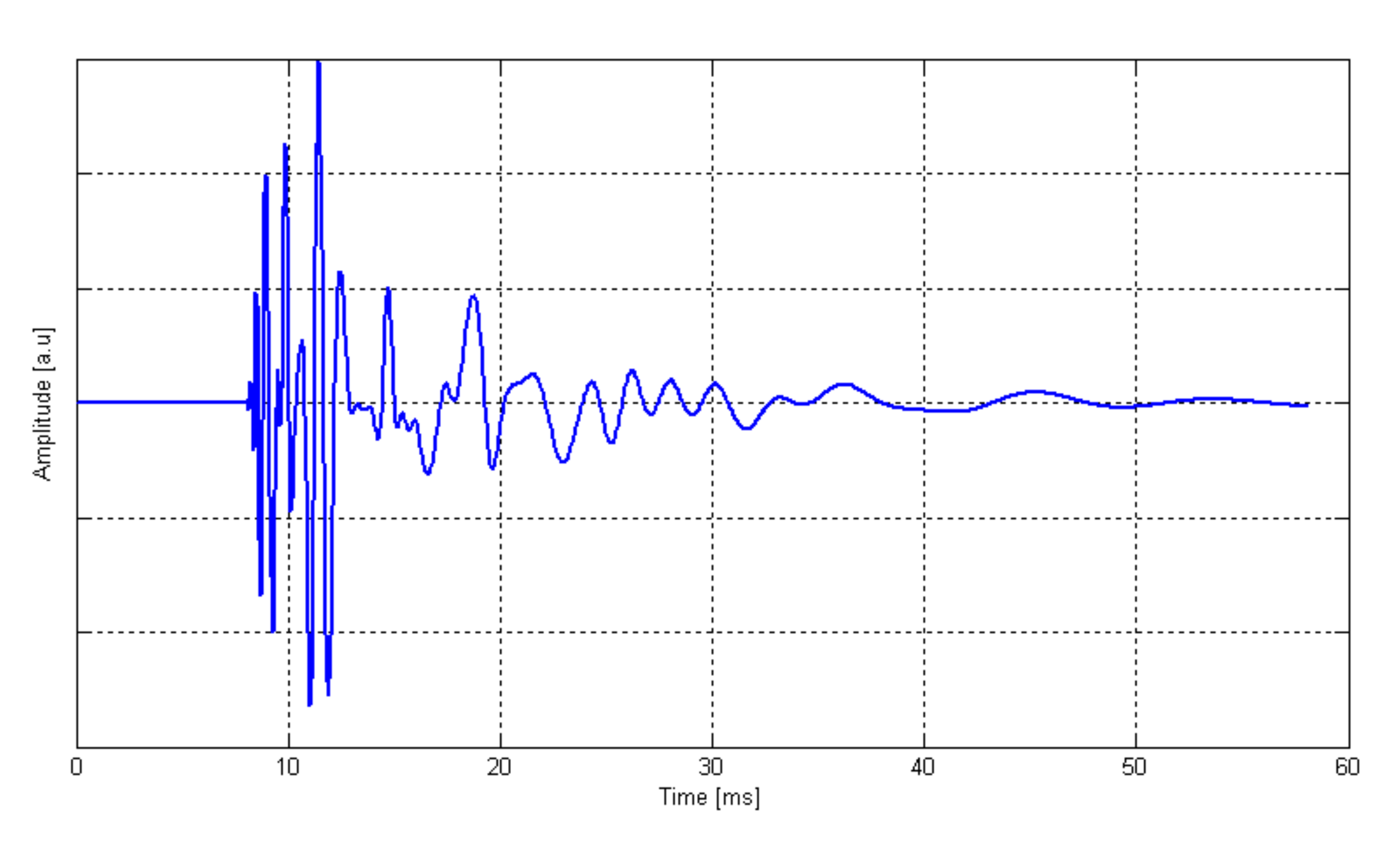}
\centering
\caption{\small{Simulated signal ($\vartheta = 10^{-5}\mathrm{s}$).}}
\label{sig_sim}
\end{figure}

Considering a rectangular room of width $l_x = 3.6 \mathrm{m}$ and length $l_y = 5.4 \mathrm{m}$, a source position at $x_e = 0.9 \mathrm{m}$, $y_e = 1.35 \mathrm{m}$ ($P7$) and a sensor position at $x_r = 0.2 \mathrm{m}$, $y_r = 5.2 \mathrm{m}$ ($C7$). The parameters for the short-term Fourier transform are a Hamming window of length of $128$ ($6.4 \mathrm{ms}$), which overlapps a segment length $126$, a fast Fourier transform length $128$, and frequency sampling of $20 \mathrm{kHz}$. The simulated signal and time frequency spectrogram are given in Figures \ref{sig_sim} and \ref{spec_demo} for a loss factor $\vartheta=10^{-5}\mathrm{s}$. The simulated time-frequency spectrogram for a loss factor of $\vartheta = 10^{-6} \mathrm{s}$ and $\vartheta = 10^{-4} \mathrm{s}$ are given in Figure \ref{spec_demo2}. As compared to the experimental spectrogram signal given in Figure \ref{P7}, the more similar spectrogram signal is given for $\vartheta = 10^{-5} \mathrm{s}$. 


%%-------------------------
\clearpage
\section{}
\label{annexeB}
In this appendix, we derive the detailed calculations to establish the relation between the propagation distance and the time of arrival of the wave packet. This relation is at the root of what is referred to as the 'perceived' propagation velocity introduced in the paper. We begin by introducing the propagation equation and related assumptions. 
The propagation equation for the displacement field in a slab writes as follows : 
\begin{eqnarray}
\rho h \frac{\partial^2 u}{\partial t^2}  + D\left(1+ \vartheta \frac{\partial }{\partial t}\right) \Delta^2 u = f \label{ep}
\end{eqnarray} 
where $f$ is the source term. Coefficients appearing in that equation are described in section \ref{TDOA}. In that appendix, we deal only with a 1D field \textit{i.e.} the field $u$ depends only on one variable $x$. Note that it would be more rigorous to derive approximation \cite{Banerjee_2004} by considering a 2D field.\\
We start by expressing (\ref{ep}) in the Fourier domain $(\omega,k)$: 
\begin{eqnarray}
\left(- \rho h \omega^2  + D\left(1- j \vartheta \omega \right) k^4\right) \hat{\tilde{u}} = \hat{\tilde{f}} \label{epf}
\end{eqnarray} 
We consider a spatio-temporal impulsion source $f$, and we assume it is separable, i.e. it may be expressed as
\begin{equation}
f(x,t) = \delta(x-x0)f(t-t0)
\end{equation}
where $\delta()$ stands for the Dirac distribution. 
In the sequel, $x_0$ and $t_0$ will be set to 0, without loss of generality. Note that for $f(t)=\delta(t)$, $u$ is the Green function of the plate. For the choice of $f$ expressed above, we get
\begin{equation}
\hat{\tilde{f}}(k, \omega)= \hat{f}(\omega)
\end{equation}
and the displacement field in the  $(\omega,k)$-domain is:
\begin{eqnarray}
 \hat{\tilde{u}}(k, \omega) = \frac{\hat{{f}}(\omega)}{- \rho h \omega^2  + D\left(1- j \vartheta \omega \right) k^4} \label{uf}
\end{eqnarray} 

\begin{figure}[!ht]
\centering
\includegraphics[width=10cm]{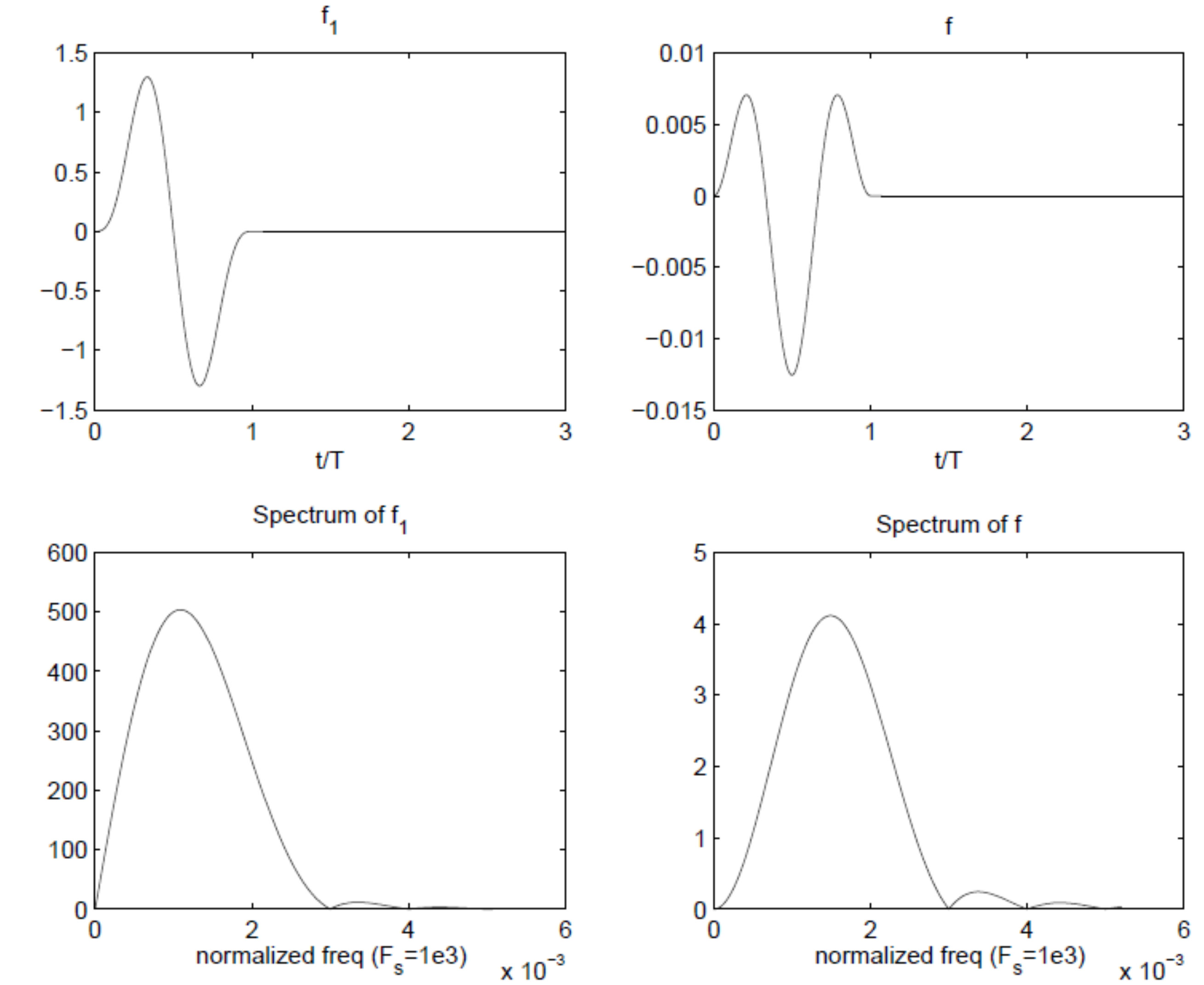}
\centering
\caption{\small{Temporal shapes and spectral content of the pulses $f1$ (left hand side) and $f$ (right hand side). }}
\label{Fig_pulse}
\end{figure}

We consider now the $(\omega,x)$-domain in which equation (\ref{eq7}) is expressed. The inverse Fourier transform (with respect to $k$) of equation (\ref{uf}) is given by:
 \begin{eqnarray}
 \hat{u}(x, \omega) &=& \int  \hat{\tilde{u}}(k, \omega) e^{j kx} dk  \nonumber \\
  &=& \int  \frac{\hat{\tilde{f}}(k, \omega)e^{j kx}dk }{- \rho h \omega^2  + D\left(1- j \vartheta \omega \right) k^4}  
   \nonumber \\
   &=& \frac{1}{\beta(\omega)}\int  \frac{ \hat{f}(\omega)e^{j kx}dk}{(k-K)(k-jK)(k+jK)(k+K)}  
  \label{uw}
\end{eqnarray} 
where $\beta(\omega)=D\left(1- j \vartheta \omega \right)$, and, where $K=\left(\frac{\rho h \omega^2}{\beta(\omega)} \right)^{1/4}$ is the pole of expression (\ref{uf}) with positive real and imaginary parts $k_R$ and $k_I$ respectively.\\
The residue theorem can be applied to evaluate expression (\ref{uw}). The upper and the lower semi-circles of radius $R$ for $x<0$ and $x>0$ respectively are considered, and the Jordan are applied on the circle parts of the domains. Finaly, we obtain the following expression of the field valid for both cases $x>0$ and $x<0$:
   \begin{eqnarray}
 \hat{u}(x,\omega) &=&  \frac{2j \pi \hat{\tilde{f}}(k, \omega)}{4\beta(\omega)K(\omega)^3}\left(e^{j K(\omega) |x|} + j e^{- K(\omega) |x|}   \right) \\
 %&=&  (1+j)^{-1}\hat{u}(\omega,0)\left(e^{j K |x|} + j e^{- K |x|}   \right) \\
 &=& U(0,\omega)\left(e^{j K(\omega) |x|} + j e^{- K(\omega) |x|}   \right)
 %&=& U(0,\omega)  \left(e^{j K |x|} + j e^{- K |x|}   \right)
\label{uw}
\end{eqnarray} 
where we define
\begin{equation}
U(0,\omega) =  \frac{2j \pi{\hat{f}}(\omega)}{4\beta(\omega)K(\omega)^3}
\end{equation}
This expression highlights the decomposition of the field in the $(\omega,x)$-domain into two exponential terms. From now on, we assume a low dissipation (\textit{i.e.} $\omega \vartheta << 1$) and a far-field context. Under these assumptions,
\begin{eqnarray}
K(\omega) & \approx & \left(\frac{\rho h \omega^2}{D} \right)^{1/4} (1+j\frac{1}{4}\vartheta \omega)\\
  &=& k_R(\omega) (1 + j\frac{1}{4}\vartheta \omega ) = k_R(\omega) + jk_I(\omega)
\end{eqnarray}
As $\frac{k_I}{k_R}\approx \frac{\vartheta\omega}{4} << 1$, we have (using $K$ instead of $K(\omega)$ for sake of readability)
\begin{eqnarray}
e^{jK|x|}+je^{-K|x|} &=& e^{jk_R|x|}e^{-k_I|x|}+je^{jk_I|x|}e^{-k_R|x|} \\
&\approx& e^{jK|x|}
\end{eqnarray}
Finaly, expression (\ref{uw}) can be approximated by:
   \begin{eqnarray}
 \hat{u}(x,\omega) \approx U(0, \omega) e^{j K |x|}  
  \label{uw2}
\end{eqnarray} 
where, using again the far field and low dissipation assumptions 
\begin{eqnarray}
U(0,\omega) \approx \frac{j \pi \hat{f}(\omega)}{2Dk_R(\omega)^3} 
\end{eqnarray}
$u(x,t)$ is obtained by computing the inverse Fourier transform wrt $\omega$ : 
\begin{eqnarray}
 u(x,t) & = & \frac{1}{2\pi} \int_{\mathbb{R}} U(0,\omega) e^{-k_I(\omega) |x|}  e^{j( k_R(\omega) |x| - \left(\omega t \right))} d\omega \\ 
& = & \frac{1}{2\pi} \int_{\mathbb{R}}  \alpha \omega^{-\frac{3}{2}} \hat{f}(\omega)e^{-\gamma\omega^{\frac{3}{2}}} e^{j\left(k_R(\omega) |x| - \omega t \right)} d\omega
\label{utx}
\end{eqnarray} 
where $\alpha = \frac{j\pi a^{3/2}}{2D}$, $\gamma=\frac{\vartheta a^{-1/2}}{4}$ and $a=\sqrt{\frac{D}{\rho h}}$.
Note that around $\omega = 0$, the integrand goes like  $\omega^{-3/2}\hat{f}(\omega)$ and cannot be integrated for any arbitrary function $f(\omega)$. A classical pulse shape \cite{Banerjee_2004} used in this framework is    
\begin{equation}
f_1(t) = \left\{ \begin{array}{l} \sin(\frac{2\pi t}{T}) - .5\sin(\frac{4\pi t}{T}) \mbox{\hspace{3mm} {\rm if} \hspace{3mm}} 0\leq t \leq T \\
				0  \mbox{\hspace{3mm} {\rm if} \hspace{3mm}}	|t| > T
				\end{array} \right.
\end{equation}
whose Fourier transform is easily obtained 
\begin{equation}
\hat{f_1}(\omega) = \frac{jT}{4\pi} e^{\frac{j\omega T}{2}} \frac{\sin(\frac{\omega T}{2})}{\left[1-\right( \frac{\omega T}{2\pi}\left)^2 \right] \left[1-\right( \frac{\omega T}{4\pi}\left)^2 \right]}
\end{equation}
which satisfies $\hat{f_1}(\omega)\propto \omega T$ when $\omega T\rightarrow 0$. One easily checks that the integral in Eq. (\ref{utx}) is now defined. However, although such an integral can be avaluated by numerical methods, fluctuations around $\omega \simeq 0$ are proportional to $\omega^{-1/2}$ do not allow an easy analytical derivation. For that latter purpose, we propose to consider  the propagation of the  perturbation associated to the first time derivative of $f_1(t)$, whose Fourier transform varies like $\omega^2$ around $\omega T\simeq 0$\footnote{Let us notice that $T$ can be arbitrarily set to very small values. A simple rescaling of the amplitude (e.g. by $1/T^2$ ) of the pulse avoids its energy to converge towards 0.  }. The shapes and spectral contents of $f_1$ and $f$ respectively are shown on figure \ref{Fig_pulse}.\\
For such an excitation,  Eq. (\ref{utx}) becomes 
\begin{equation}
u(x,t) \simeq \frac{1}{2\pi} \int_{\mathbb{R}}  \alpha \omega^{+\frac{1}{2}} e^{-\gamma\omega^{\frac{3}{2}}} e^{j\left(k_R(\omega) |x| - \omega t \right)} d\omega
\label{utx2}
\end{equation}
The regularity of the integrand in Eq. (\ref{utx2}) allows to rely on the stationary phase method to evaluate $u(x,t)$.\\
The stationary phase condition leads to 
\begin{equation}
\omega_0 = \frac{1}{4a}\frac{x^2}{t^2}
\label{omega0}
\end{equation}
and the enveloppe of $u(x,t)$ satisfies
\begin{equation}
A(x,t) \propto \frac{|\alpha| \omega_0^{1/2}e^{-\gamma \omega_0^{3/2}|x|}}{\sqrt{x|k"(\omega_0)|}} \\
	\propto 2 |\alpha| a^{1/4}\omega_0^{5/4}x^{-1/2}e^{{-\gamma \omega_0^{3/2}|x|}}
\end{equation}
At a given time instant, the maximum of the perturbation is located at $d$ satisfying $\frac{\partial A}{\partial x}(d) =0$.
using the expression of $\omega_0$ from Eq. (\ref{omega0}),  leads to the condition  
\begin{equation}
\frac{\partial}{\partial x} \left( 2^{-3/2}|\alpha| a x^2 t^{-5/2} e^{-\frac{\gamma}{8a^{3/2}}x^4t^{-3}}\right) = 0 
\end{equation}
or equivalently 
\begin{equation}
1- \frac{\vartheta d^4}{16a^2t^3} = 0.
\end{equation}
%---
\clearpage
\end{appendix}
%--------------------
\section*{Acknowledgements}
This study was supported in part by the Atomic Energy Commission (CEA) Grenoble, by the National Polytechnical Institute (INPG), and by the National Engineering School of Tunis (ENIT) under a cooperative agreement.
%---------------------------- 
%\bibliographystyle{model1-num-names}
%\bibliography{Bibliog}
%\input{xarchive.bbl}

\section*{}

\clearpage
\listoffigures
\end{document}